\documentclass{article}

\usepackage{amsmath}
\usepackage{graphicx}
\usepackage{grffile} 
\usepackage{graphics}
\usepackage[usenames,dvipsnames]{color}
\usepackage[svgnames]{xcolor}
\usepackage{colortbl}
\usepackage{fancybox}
\usepackage{hyperref}
\usepackage{cleveref}
\usepackage{fancyvrb}
\usepackage{amssymb}
\usepackage{amsfonts}

\usepackage{amsthm}

\usepackage{calc}
\usepackage{float}
\usepackage[inline]{enumitem}
\usepackage{xstring}
\usepackage{tabu}
\usepackage{hyperref}
\hypersetup{
  hidelinks,
  pdfborder={0 0 0},
  colorlinks=false,
  urlbordercolor={0 0 0}
}

\newif\ifsmallFigures 
\newif\ifconf 
\newif\ifconfTfpie 

\conffalse
\confTfpiefalse

\ifconf
   \smallFigurestrue
\else
   \smallFiguresfalse
\fi

\theoremstyle{definition} \newtheorem{definition}{Definition}

\theoremstyle{definition} \newtheorem{example}{Example}
\theoremstyle{definition} 
\theoremstyle{definition} 
\theoremstyle{definition}

\newcommand{\todo}[1]{}
\newcommand{\answer}[1]{}
\newcommand{\coulddo}[1]{}
\newcommand{\shoulddo}[1]{}
\newcommand{\mustdo}[1]{} 
\newcommand{\todoforchide}[1]{}

\newcommand{\done}[1]{} 
\newcommand{\forNextIncrement}[1]{} 
\newcommand{\proofread}[2]{} 

\newcommand{\deprecated}[1]{}

\newcommand{\potentialTaskForAlle}[1]{}

\newcommand{\omitForInTestLatexBuilt}[1]{#1}

\newcommand{\coreng}[1]{\correctEnglish{#1}} 
\newcommand{\correctEnglish}[1]{}

\definecolor{LightGray}{gray}{0.9}

\newcounter{questioncount}

\potentialTaskForAlle{\coulddo{&y2016.01.07.13:04:41& how to create a subcounter for when a newenvironment can be used in a nested way?}}

\newcommand{\comm}[1]{}
\newcommand{\talkingscriptgeneral}[1]{}
\newcommand{\talkingscriptmgmag}[1]{}

\shoulddo{&y2015.12.25.19:07:56& make it so that secref, changes to `part' or `paragraph' etc. depending on the type of part of the work is beign referred to.}

\newcommand{\comment}[1]{}
\newcommand{\commentByAlody}[1]{}
\newcommand{\commentByDouweSchulte}[1]{}
\newcommand{\adoptedCommentByAlody}[1]{}
\newcommand{\deprecatedCommentByAlody}[1]{}
\newcommand{\commentByChide}[1]{}
\newcommand{\clog}[1]{} 
\newcommand{\decision}[1]{} 

\newcommand{\behaviouralStructure}[1]{}
\newcommand{\bs}[1]{\behaviouralStructure{#1}}

\newcommand{\obsolete}[1]{}
\newcommand{\omitted}[1]{} 

{%
\begin{figure}%
\begin{Sbox}%
\begin{minipage}[t]{\textwidth}%
}%
{%
\end{minipage}%
\end{Sbox}%
\doublebox{\TheSbox}%
\end{figure}%
}

{%
\begin{figure}[H]%
\begin{Sbox}%
\begin{minipage}[t]{\textwidth}%
}%
{%
\end{minipage}%
\end{Sbox}%
\doublebox{\TheSbox}%
\end{figure}%
}

\newcommand{\commforchide}[1]{}

\newcommand{\commforjj}[1]{}

\newcommand{\feedbackjj}[1]{}

\newcommand{\steno}[1]{}

\newcommand{\covsteno}[1]{} 

\newcommand{\cons}[1]{} 

\newcommand{\omitinversionwithoutcomments}[1]{}

\newcommand{\notfinishedyet}[2]{} 
\newcommand{\dontshow}[1]{\textsf{[At this place there is a text fragment that is not shown, for example because it is still work in progress.]}}
\newcommand{\sketch}[1]{} 

\newcommand{\rewrite}[2]{} 

\newcommand{\sourcecode}[1]{\texttt{#1}}
\newcommand{\sourcecodeQuoted}[1]{`\texttt{#1}'}
\newcommand{\scq}[1]{`\texttt{#1}'}

\DefineVerbatimEnvironment%
{sourcecodeEnv}{Verbatim}
{gobble=0,numbers=left,numbersep=2mm,
frame=lines,framerule=0.8mm,fontsize=\small}

\newcommand{\urlref}[1]{\footnote{\url{#1}}}

\newcommand{\fltu}[1]{\firstLetterToUppercase{#1}}
\newcommand{\firstLetterToUppercase}[1]%
{\StrLeft{#1}{1}[\firstletter]%
\StrGobbleLeft{#1}{1}[\rest]%
\expandafter\MakeUppercase\expandafter{\firstletter}\rest%
}

\done{{| y2018_m02_d10 |} change newterm to newtermtest (I already tested it successfully), and change the applications of it throughout the book.}
\newcommand{\newterm}[1]{\emph{\glsdef{#1}}\marginpar{\emph{\textsf{\small{\gls{#1}}}}}}

\newcommand{\voicerec}[1]{} 

\newenvironment{visupolfigure}%
{%
\begin{figure}[H]%
\begin{minipage}[t]{0.99\columnwidth}%
}%
{%
\end{minipage}%
\end{figure}%
}

{%
\begin{figure}[H]%
}%
{%
\end{figure}%
}

\renewcommand{\newterm}[1]{\emph{#1}}

\newcommand{\adtpolyOne}{Algepoly1} 
\newcommand{\adtPolyOneTooFlexible}{Algepoly1$^+$} 
\newcommand{\fplE}{functional programming language}
\newcommand{\fpl}{FPL}
\newcommand{\trans}{\mathit{trans}}

\newcommand{\madawipolA}{Madawipol-$\alpha$}
\newcommand{\MadawipolA}{Madawipol-$\alpha$}

\bs{{| y2016_m06_d23_h14_m31_s17 |} don't give it the same name as frapoly, because frapoly is bit `biger' than what is maramafied (there are no algebraic data structure definitions)}

\newcommand{\madawipolB}{Madawipol-$\beta$}

\newcommand{\alignmentSquare}{alignment square}

\newcommand{\jointForm}{joint-form} 
\newcommand{\femaleJntForm}{female joint-form}
\newcommand{\FemaleJntForm}{Female joint-form}
\newcommand{\maleJntForm}{male joint-form}
\newcommand{\MaleJntForm}{Male joint-form}

\newcommand{\verticalJointSize}{vertical joint size}

\newcommand{\FverticalJointSize}{\mathit{vJntSz}}

\newcommand{\typeCons}{type-constructor}
\newcommand{\typeParameter}{type-parameter}
\newcommand{\typeConsForm}{type-constructor form}
\newcommand{\TypeConsForm}{Type-constructor form}
\newcommand{\rigidPart}{rigid part}

\newcommand{\TypeForm}{Type-form}
\newcommand{\typeForm}{type-form}
\newcommand{\ADT}{algebraic data type}
\newcommand{\ADTD}{algebraic data type definition}

\newcommand{\FsetADTD}{\mathit{ADTDset}}
\newcommand{\typeConsMap}{type-constructor mapping}
\newcommand{\FtransConfig}{\mathit{tConf}}

\newcommand{\polymorphicSubspace}{polymorphic subspace}
\newcommand{\polymorphicSurface}{polymorphic surface}
\newcommand{\PolymorphicSurface}{Polymorphic surface}
\newcommand{\polySurfConstantFORM}{\mathit{polSurf}}
\newcommand{\femaleBottomRegion}{female bottom region}

\newcommand{\mimickingPolSurf}{mimicking}
\newcommand{\MimickingPolSurf}{Mimicking}

\newcommand{\differentiatedCXTpS}{differentiated}

\newcommand{\undifferentiatedCXTpS}{undifferentiated}

\newcommand{\differentiatedRegion}{differentiated region}

\newcommand{\mimickingRegion}{mimicking region}

\newcommand{\undifferentiatedRegion}{undifferentiated region}

\newcommand{\freeRegion}{free region}

\newcommand{\mconstructor}{M-constructor}

\newcommand{\mco}{M-construct}
\newcommand{\constrBlock}{constructor block}
\newcommand{\constrBlockMap}{\constrBlock\ mapping}

\newcommand{\constrArgsMap}{constructor argument-location mapping}

\newcommand{\transConfig}{translation configuration}

\newcommand{\tcArgumentTransformation}{\typeCons\ argument transformation}

\newcommand{\Fads}{\mathit{ads}} 
\newcommand{\adsE}{algebraic data structure}
\newcommand{\ads}{ADS}
\newcommand{\Mads}{M-ADS}
\newcommand{\MadsE}{M-algebraic data structure}

\newcommand{\incompleteAds}{unfinished \ads}

\newcommand{\constrResultTypeMap}{constructor's result-type location mapping}

\newcommand{\Funjoinable}{\mathsf{unjoinable}}

\mustdo{check term with the literature}

\newcommand{\FunifiableWith}{=_{u}}
\newcommand{\FnotUnifiableWith}{\neq_{u}}

\mustdo{{| y2016_m11_d27_h20_m16_s10 |} change such that also variables can be passed (so not only \sourcecode, e.g. constr instead of Cons)}

\newcommand{\polyBlocks}{polymorphic blocks}
\newcommand{\polyBlocksA}{PB}

\newcommand{\Mcons}[1]{M-\sourcecode{#1}}
\newcommand{\Mtype}[1]{M-`\sourcecode{#1}'}

\newcommand{\MflexCFORM}{\mathit{MFC}}

\newcommand{\emptySpot}{\sourcecode{\_}}

\begin{document}

\omitted
{ Rules to follow when writing this article.
   { Fixed terminology:

      o visual programming construct = VPC <g7 &y2013.03.14.23:26:57& idea: instantly comprehensible visual programming constructs icvpc's or so>
      o maramafied constructor
      o a ball which holds
      o Always use one of: `visual blocks' (for things that are actually blocks), or `\mcos' (for anything visual), TODO finish. And not, for example: 
      o Use `polymorphism' instead of `polymorphy'.
      { log
        [{| y2018_m02_d20_h12_m35_s36 |} is it programming with polymorphism or polymorphy? Or are both possible. Did some research on the Web, and it seems polymorphism is used more often.]
      }
   }
}


\mathchardef\mhyphen="2D

\newcommand{\scaleFactor}{0.1}
\newcommand{\scaleFactorA}{0.1}
\newcommand{\scaleFactorB}{0.1}
\newcommand{\scaleFactorC}{0.1}
\newcommand{\scaleFactorD}{0.1}
\newcommand{\localcaption}{[??]}
\newcommand{\figA}{[??]}
\newcommand{\figB}{[??]}


\newcommand{\titleText}{Truly Visual Polymorphic Algebraic Data Structures through \emph{Maramafication}}
\title{\titleText}


\author{Chide Groenouwe, Jesse Nortier \& John-Jules Meyer\\
Universiteit Utrecht\\
Information and Computing Sciences\\
Utrecht, the Netherlands\\
\{c.n.groenouwe$\mid$j.j.c.meyer\}@uu.nl\ jessenortier@gmail.com
}

%




\maketitle

\begin{abstract}
This paper presents a so-called \emph{maramafication} of an essential part of \fplE s such as Haskell or Clean: the construction of fully polymorphic well-typed \adsE s based on type definitions with at most one type parameter.\shoulddo{{| y2018_m03_d15_h19_m36_s31 |} tell somewhat more about what this constitutes: that well-typedness is enforced completely mechanically, without requiring any explanation.} As such, this work extends our previous work, in which only a very limited form of polymorphism was present \cite{groenouwe2017}. Maramafication means the design of visual `twins' of existing programming constructs using spatial metaphors rooted in common sense or inborn spatial intuition, to achieve self-explanatoriness. This is, among others, useful to considerably reduce the gap between programmers and non-programmers in the creation of programs, for educational purposes, for inclusion of non-typical programmers and for invoking enthusiasm among non-programmers.
\end{abstract}


\mustdo{{| y2018_m02_d20_h12_m45_s08 |} rephrase it such that the formulation is not identical to previously submitted papers.}

\section{Introduction}

It would be highly beneficial if non-programmers could co-program software applications. The Marama-paradigm\todo{y2016_m01_d27_h16_m29_s54 other name than marama needed - it has to do with the way semantics is realised.}, as introduced in previous work\todo{cite{}}, is a paradigm that is intended to considerably reduce the gap between programmers and non-programmers. The basis of the Marama-paradigm consists of designing `twins' of programming constructs using metaphors rooted in common sense or inborn intuition, making the constructs almost entirely self-explanatory. This work has coined the term \newterm{internal semantics} for this purpose: the semantics of constructs is evident without an external definition. This may lower the threshold for non-programmer participation.\shoulddo{{| y2018_m05_d16_h15_m56_s14 |} not only internal, also learning laws by example should be included.} Other application areas of maramafication include education and fostering an inclusive society. For example, some dyslexics may have a latent talent for programming that never surfaces in a world dominated by textual programming languages\mustdo{ \cite{}}. Some authors have argued that these people would be a great asset to the programmer's community, because they believe there is a correlation between dyslexia and the capability to think more creatively than the average.\mustdo{TODO include citation from Communications of ACM.}

\coulddo{{| y2018_m05_d16_h16_m45_s54 |} finish: Indeed, preleminary experimentationExperimentation with maramafied constructs indeed exposed people with poor prog, to perform unexpected good. }

This work coins the term \newterm{maramafication} for this design process. In other words, if such a twin has been designed for a language construct from for example Clean \cite{Brus87} or Haskell \cite{Hudak92}, it has been `maramafied'.  In the remainder of this article, the prefix `M-' indicates `maramafied', for example, an \mconstructor, is a maramafied constructor.

Related work, among others in the field of visual general purpose programming languages, to the best of our knowledge, does not truly and fully employ this principle of internal semantics. Note that visualisation itself is not equal to having an internal semantics. Many `visual' constructs occurring in visual programming languages are not self-explanatory. What is more, most of such languages do not even visualise most of the programming constructs. They are still `contaminated' with many constructs of a textual nature. Typically, in such programming languages the truly `visual' part consists of boxes connected with arrows, while the content of the boxes contains much textually expressed programming code, such as textually defined typing information and data-structures. Hence, these languages require the user to still largely move within the original not-so-accessible textual programming paradigm. The hardest parts to truly capture with internal semantics are, possibly unconsciously, left in a textual form. In this article, however, we cover such a hard and non-trivial part.\mustdo{TODO citations to be included}

This paper focusses on a fragment of the `maramafication-challenge': it presents a new way to (truly) visually represent fully polymorphic \adsE s (\ads s) as they occur in modern functional programming languages, and does so in line with the aforementioned paradigm. The design is modular: it can be adopted straightforwardly into any visual functional programming language.\shoulddo{{| y2016_m02_d25_h15_m59_s54 |} A bit vague like this, add that it can be combined with visualisations of other programming constructs without a problem.}\coulddo{{| y2016_m01_d27_h17_m33_s41 |} give examples! Also as a kind of invitation to the developers of those languages.}\shoulddo{{| y2016_m02_d25_h16_m03_s22 |} mention that you cover `some forms of' polymorphy, to prevent making an overstatement. For example, I believe I do not cover polymorphic function types.}

\fltu{\ads}s are designed in such a way that type consistency is entirely forced by the form of the \mco s (maramafied constructs). In other words, a user of these constructs cannot create ill-typed values, simply because the `pieces will not fit'. In this sense, the approach in this paper is genuinely visual: the semantics of the visual blocks is embodied by their visual structure and spatial manipulation options, and does not require a definition by textual or spoken means. Hence, a beginner using the \mco s can find out how to program with them, without any prior textual or spoken explanation about how these constructs work.\shoulddo{&y2016.02.24.19:53:54& Most visual language are not `truly' visual in the sense that they use other. Also provide a clear example and counter-example.}

Another way to phrase it, is that the semantics of the visualisation of polymorphy and datastructures proposed in this paper solely relies on shared human intuition for manipulation of 3D objects.

In this article, the term \newterm{spatial necessity} is coined for the aforementioned property of the visual designs, the property that given the laws of mechanics (as far as they are intuitively understood by the majority of humans) it is only possible to construct something that is correct.\mustdo{{| y2016_m01_d27_h17_m40_s46 |} brief intro in how it works.} An example of such a widely shared intuition on which the spatial necessity design paradigm can rely, is that most people from an already very young age will predict that a ball that is held in the air, and then let lose, will move downward.

The design covers \ads s based on \ADTD s with at most one \typeParameter\ (but is easily extensible with any number of \typeParameter s), can cope with polymorphic constructors, and can classify a given \ads\ polymorphically (through `typing statements'). 

\Cref{SEC__Textual-Languages} presents the textual languages used to show the textual equivalences of the \mco s. \Cref{SEC__MadawipolB-by-Example} introduces a predecessor to this work: \madawipolA: a maramafication of \mconstructor s that is already sufficiently expressive to deal adequately with recursively defined types \cite{groenouwe2017}. However, \madawipolA\ does not yet support full parametric polymorphism. \Cref{SEC__MadawipolB-by-Example} introduces \madawipolB, an adaptation of \madawipolA\ that adds parametric polymorphism. \Cref{SEC__Definition-MadawipolB} provides definitions of essential aspects of \madawipolA.\shoulddo{{| y2018_m08_d25_h02_m37_s12 |} finish the overview.}

It is important to note that this paper explains the constructs primarily from the perspective of a programming language designer, the targeted readers of this article. The reader has to keep in mind that the targeted users of the language, however, will learn how the constructs work by a playful and wordless exposure to maramafied \ads s or parts thereof only. No textual explanations will be provided. Moreover, the users of the language will use an interactive 3D-editor. The \shoulddo{`noodgedwongen'} static representation in this article, therefore, required some alternative ways to represent aspects of the design. It is important that the reader does not confuse these with the actual, and more `intuitive', way the language will occur to the user.\shoulddo{mention some examples, or refer to them.}

\section{Textual languages}\label{SEC__Textual-Languages}

\shoulddo{{| y2018_m03_d15_h19_m41_s59 |} perhaps it is a good idea to have at the least two ADTs with a type parameter, for example List and Tuple, to emphasise the generality of the design.}
This section\shoulddo{does not work: \crefname{SEC__Examples__Textual-Language}} presents the textual languages that are used in this article to show the textual equivalences of the \mco s presented in this paper. Because the \mco s covered in this paper only deal with a fragment of a modern functional programming language (\fpl) such as Clean or Haskell, this article also defines a textual language that is (isomorphic to) the relevant fragment of such an \fpl. The language is called \adtpolyOne. We suffice with introducing the textual languages by example, their semantics as identical to the corresponding parts of Clean.\omitted{ (A formal specification is to be found in \cref{SEC__Definition}.)}

\subsection{\adtpolyOne}

\begin{example}[Algebraic Data Type Definitions]\label{EX_--_-ADTDs}
The following \emph{Algebraic Data Type Definitions} define a number of algebraic data types:
\begin{sourcecodeEnv}
::WeekendDay = Sat  | Sun
::Bool       = True | False
::Colour     = Red | Blue | Green
::List a     = Cons a (List a) | Nil
\end{sourcecodeEnv}
\end{example}

Note that this paper does not cover a visual counterpart to \ADTD s. However, it is important to include them in the textual language for explanatory and definitory purposes.

\begin{example}[\ads s]
The following is a comma-separated list of \ads s:
\begin{sourcecodeEnv}
True, Blue, Cons True Nil, Red,
Cons True (Cons True Nil).
\end{sourcecodeEnv}
\end{example}

In this paper it is important to be able to talk about \adsE s that are still under construction.  For this purpose, we add the following construction.

\begin{example}[\incompleteAds s]\label{EX_--_-Unfinished-ADS}
In \incompleteAds s not all argument positions of constructors have been supplied with arguments, for example:
\begin{sourcecodeEnv}
Cons _ (Cons True Nil), Cons Red (Cons Green _)
\end{sourcecodeEnv}
The `\emptySpot' stands for an `empty spot'. Such an `\emptySpot' may only occur at the argument-position of a constructor application.
\end{example}

This paper also deals with polymorphic constructors. These normally do not occur in languages such as Clean as `stand-alone' constructors. To be able to talk about them in isolation, \adtpolyOne\ contains the following constructs.
\begin{example}
The following statement:
\begin{sourcecodeEnv}
Cons:[List Bool]
\end{sourcecodeEnv}
is in a spoken phrase expressed as ``the \sourcecodeQuoted{Cons} of \sourcecodeQuoted{List Bool}.''. Such a statement only makes sense in the context of an \ADTD, in which the constructor has been defined. Lets assume that this \ADTD\ is the one in \cref{EX_--_-ADTDs}. Then this statement expresses that \sourcecodeQuoted{Cons} is the constructor that one would get by substituting \sourcecodeQuoted{Bool} for the \typeParameter\ in the given \ADTD.\omitted{cannot be written down yet, because notation has not yet been introduced} In this case that would be the \sourcecodeQuoted{Cons} with the following type: \sourcecodeQuoted{Bool (List Bool) -> List Bool}. Another way to phrase it, is that it is the instance of \sourcecodeQuoted{Cons} with result-type \sourcecodeQuoted{List Bool}.

\clog{
[[{| y2018_m05_d09_h14_m04_s12 |} No, it *is* correct after all. With it I want to indicate the Cons that belongs to the type definition of List a.]

[{| y2018_m05_d09_h14_m02_s05 |} this is not correct! What I mean to say is that it is the Cons with result-type List ( List a). The type indicated in this statement is the value of the \typeParameter\ of the ADTD definition.]
}
Another instance of \sourcecodeQuoted{Cons} is 
\begin{sourcecodeEnv}
Cons:[List (List a)]
\end{sourcecodeEnv}
``the \sourcecodeQuoted{Cons} of \href{https://drive.google.com/open?id=1hOt8s2Igb5Q7_yNQgrEWv1AxJuQR4GjU}{\sourcecodeQuoted{List (List a)}}.'' This instance of \sourcecodeQuoted{Cons} is, clearly, polymorphic.
\end{example}

\subsection{\adtPolyOneTooFlexible}

For didactic purposes, it turned out to be useful to introduce a slight extension of the language \adtpolyOne: \adtPolyOneTooFlexible\ (the `+' stands for `too flexible'). It allows constructors to have any type, including the most general polymorphic result-type \sourcecodeQuoted{a} and non-polymorphic argument-types, or no results or arguments at all. (Hence, they are reminiscent of generalised algebraic data types \cite{Vytiniotis2006}.)  Note that the language itself is not intended for use in a real language. It is `too flexible' and would lead to type-technical and semantical difficulties. It serves as a way to create `minimal examples' to demonstrate certain behaviours of \mconstructor s. Note that the notation of the type is written down in the opposite order of what is commonly used, which one can witness by the reversed direction of the arrow. The result type is written down first, then the arrow follows, and finally the argument types are provided. This has been done for convenience: the result and the arguments are now in the same order as they appear when the constructor is applied, both in the textual version as in the maramafied version.

The following defines \adtPolyOneTooFlexible\ by means of examples.

\begin{example}\label{EXAMPLE_--_AlgepolyOnePlus}
The following definition:
\begin{sourcecodeEnv}
Red: Colour <-
\end{sourcecodeEnv}
defines a constructor \sourcecodeQuoted{Red} that has no arguments and result type \sourcecodeQuoted{Colour}.\omitted{ (To be correct: it has the `information-less' unit type as its argument, a trick to say that in a certain semantical sense does not take arguments.)}
\begin{sourcecodeEnv}
SimpleFemCons: <- Colour
\end{sourcecodeEnv}
\done{{| y2018_m05_d15_h18_m16_s16 |} how to write down a constructor without result-type in an idiomatic way?}
defines a constructor \sourcecodeQuoted{SimpleFemCons} that has one argument with type \sourcecodeQuoted{Colour}, and that does not produce a result.

\begin{sourcecodeEnv}
FlexiCons: a <- a
\end{sourcecodeEnv}
defines a constructor with one argument with the most general polymorphic type \sourcecodeQuoted{a} and a result with the same type. More examples:
\begin{sourcecodeEnv}
PolyCons: SimpleType a <- a
SimplePairCons: <- a a
\end{sourcecodeEnv}
\end{example}

\section{\madawipolA\ by example}\label{SEC__MadawipolA-by-Example}

\MadawipolA, created in our previous work, is a maramafication that covers \ads s based on \ADTD s with at most one type parameter, and can cope with a limited form of polymorphic constructors: those without arguments \cite{groenouwe2017}. This section suffices with defining \MadawipolA\ by example, among other due to space limitations, but also because the successor \madawipolB\ is the focus of this paper. For an elaborate formal definition of \madawipolA, see \cite{groenouwe2017}. Throughout the online version of this document, clicking any image shows an online high resolution version of it, which can be zoomed into. The reader is strongly recommended to follow these links to investigate the figures in much more detail. The same holds for many textual descriptions. For off-line readers  there is an additional `appendix'-document that contains an enlarged version of each figure, either to print or to store locally. It is available from the same location as this article.\mustdo{Add download link}

\coulddo{
o {| y2016_m08_d24_h13_m16_s22 |} nice would be an instance Pair Bool WeekendDay (an instance of ::PairType a b = Pair a b, because that would constitute a minimal example without polymorphism or nestable constructors (such as list). However, it is a pity that the male joint of PairType shows a more complex structure (Even multiple parameter joint!).
o {| y2016_m08_d24_h13_m21_s08 |} so, perhaps it is better to show the marafication of true <: Bool
}

\Cref{FIG__Atomic-ADSs} shows examples of the simplest possible \ads s: atomic \ads s.  In particular, note the form of the joints. The values of type \sourcecodeQuoted{Bool} have the same \newterm{\jointForm}, while \sourcecodeQuoted{Sat} has another. Every \typeCons\ has a unique form associated with it, its \newterm{\typeConsForm}. In this case, \sourcecodeQuoted{Bool} corresponds with the triangle in the figure, and \sourcecodeQuoted{WeekendDay} with the pentagon in the figure. (The \typeConsForm s have to comply with some additional conditions, these are covered in \cite{groenouwe2017}.\mustdo{{| y2018_m08_d25_h11_m50_s04 |} and refer to madawipolB in this paper.}) The square form around them is not explicitly part of the type.  It is the \newterm{\alignmentSquare}, needed to align joints correctly when fitting them together.\mustdo{add reference to the precise treatment in madawipolb.}
\begin{visupolfigure}
\omitForInTestLatexBuilt{
\renewcommand{\scaleFactor}{0.30}
\vspace{3mm}
\centering
\tabulinesep=\tabcolsep
\begin{tabu} to \textwidth {|X[1,c,m]|X[1,c,m]|X[1,c,m]|}
    \hline
\href{https://drive.google.com/open?id=1YvTWr2Q6bVV5ik9O0fLrQJfmVAWNPDLy}{\includegraphics[width=\scaleFactor\textwidth]{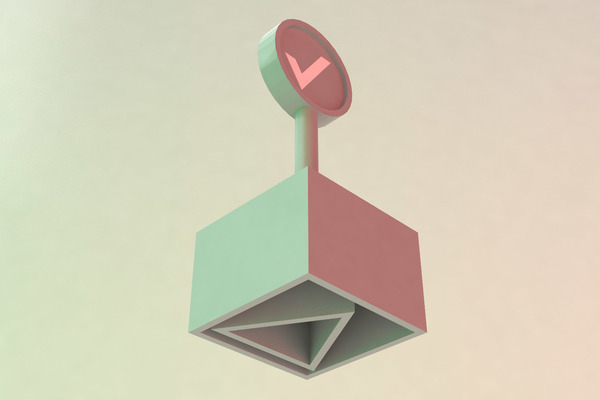}} &%
\href{https://drive.google.com/open?id=10qs1IEybKUtNg3_Lgu9rvRuKjPs4LSkT}{\includegraphics[width=\scaleFactor\textwidth]{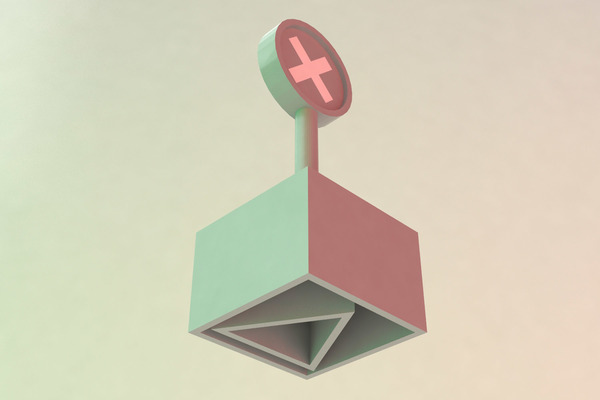}} &%
\href{https://drive.google.com/open?id=1EKlhUfUilL2UQe5LGqZta1YWMYIYBrte}{\includegraphics[width=\scaleFactor\textwidth]{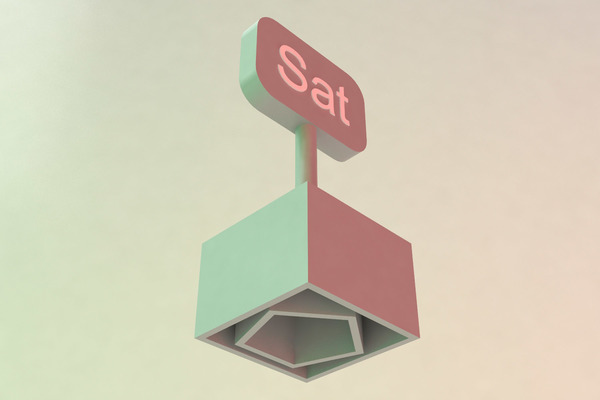}} \\
\sourcecodeQuoted{True} & \sourcecodeQuoted{False} & \sourcecodeQuoted{Sat}\\
\hline
\end{tabu}
}
\caption{Atomic \adsE s. Important: all images are clickable and lead to zoomable high resolution versions.}\label{FIG__Atomic-ADSs}
\end{visupolfigure}

\mustdo{{| y2018_m08_d25_h03_m02_s48 |} somewhere explain that the definition of madawipolA is inductive/compositional (translation), why that is powerful, and that it is not trivial.}
To provide examples of molecular \ads s (\ads s with constructors that take arguments), \cref{FIG__Molecular-ADSs__Several-mconstructors} introduces all \mconstructor s that belong to the type \sourcecodeQuoted{List Bool} and \sourcecodeQuoted{List (List Bool)}. Note that \madawipolA\ does not yet support polymorphism for constructors with arguments. Therefore, it can only provide non-polymorphic instantiations of \sourcecodeQuoted{Cons}. \Mcons{Nil}, however, is polymorphic. (Note that ``\Mcons{Nil}'' is an example of the notation of specific \mconstructor s.) Hence, it can be shared among the different instantiations of \sourcecodeQuoted{Cons}. This explains why there is only one \Mcons{Nil} in \cref{FIG__Molecular-ADSs__Several-mconstructors}.

Note the forms of the three joints of \Mcons{Cons:[List Bool]}. It has one male joint, that corresponds with the result-type \sourcecodeQuoted{List Bool}, and two female joints, one corresponding to the type \sourcecodeQuoted{Bool}, and the other to \sourcecodeQuoted{List Bool}. The circle corresponds with the \typeCons\ \sourcecodeQuoted{List}. It is an essential aspect of the design that \jointForm s that correspond with a complex type, such as \sourcecodeQuoted{List Bool}, are compositionally related to their textual form. The 2D projection of the \jointForm\ along a line perpendicular to the bottom of the joint, is called the \typeForm. I.e.~it is the 2D form one sees when viewing a joint along a line of sight that is perpendicular to the bottom of the joint. This is an adequate abstraction. After all, if one assumes that male and female joints have matching heights and depths, the only aspect that determines whether they fit is the \typeForm. The textual type can be read from the \typeForm\ by starting with the outermost form (skipping the \alignmentSquare) and then walking one's way to the adjacent one, all the way down to the center form, while reading out loud the \typeCons\ that is associated with each form. The reader is invited to verify that all joints of \Mcons{Cons:[List (List Bool)]} also comply with this structure. A polymorphic type, such as \sourcecodeQuoted{List a}, is created by leaving the space within the innermost \typeConsForm\ empty. The joint of \Mcons{Nil} is an example.

\mustdo{also add explanation of Nil, as example of how to deal with polymorphism.}

Also note the form of the three joints of \Mcons{Cons:[List (List Bool)]}. 
\begin{visupolfigure}
\omitForInTestLatexBuilt{
\renewcommand{\scaleFactor}{0.3}
\renewcommand{\scaleFactorB}{0.3}
\centering
\tabulinesep=\tabcolsep
\begin{tabu} to \textwidth {|X[1,c,m]X[1,c,m]|X[1,c,m]|}
\hline
\href{https://drive.google.com/open?id=1kOhCIDm5tvOrVrj5tryAg34ThRjrSOV0}{\includegraphics[width=\scaleFactor\textwidth]{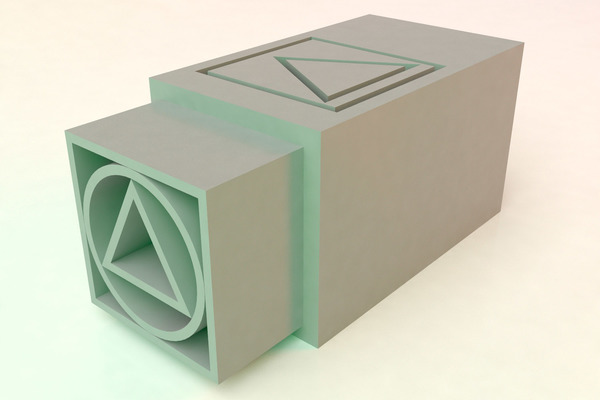}} &%
\href{https://drive.google.com/open?id=1C1xuw2OOYOF-gzBv4zDUj3T4X0pKtMUd}{\includegraphics[width=\scaleFactor\textwidth]{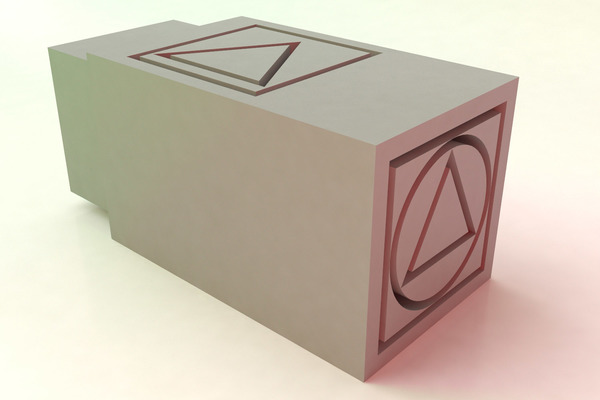}} &%
\href{https://drive.google.com/open?id=1Pxn68oMLNCFmtOiQcczzwaWZzEkbubco}{\includegraphics[width=\scaleFactor\textwidth]{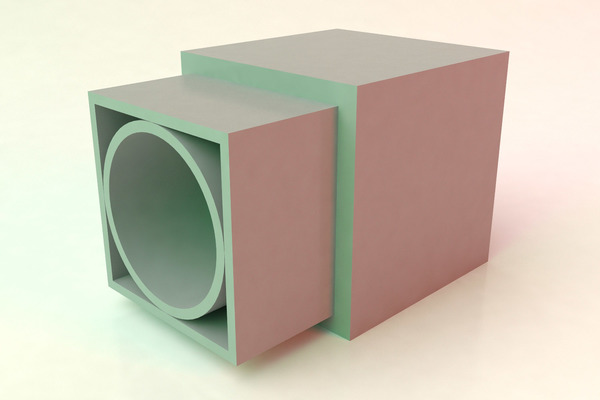}} \\
\href{https://drive.google.com/open?id=1kOhCIDm5tvOrVrj5tryAg34ThRjrSOV0}{\includegraphics[width=\scaleFactor\textwidth]{figures/resized/FIG__Molecular_ADSs__Cons-List-Bool_Perspective1.jpg}} &%
\href{https://drive.google.com/open?id=1C1xuw2OOYOF-gzBv4zDUj3T4X0pKtMUd}{\includegraphics[width=\scaleFactor\textwidth]{figures/resized/FIG__Molecular_ADSs__Cons-List-Bool_Perspective2.jpg}} &%
\href{https://drive.google.com/open?id=1Pxn68oMLNCFmtOiQcczzwaWZzEkbubco}{\includegraphics[width=\scaleFactor\textwidth]{figures/resized/FIG__Molecular_ADSs__Nil_Perspective.jpg}} \\
\multicolumn{2}{|c|}{\sourcecodeQuoted{Cons:[List Bool]} (2 perspectives)} & \sourcecodeQuoted{Nil:[List a]} \\
\hline
\href{https://drive.google.com/open?id=1P7_cbktIyYjK03EZtNR9QsQiuQ42HWRp}{\includegraphics[width=\scaleFactor\textwidth]{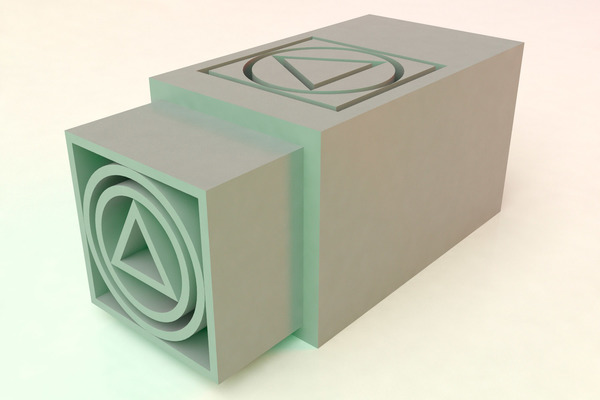}} &%
\href{https://drive.google.com/open?id=1wxO_pNx2Bonql979P-OtcULTC5n2lGJj}{\includegraphics[width=\scaleFactor\textwidth]{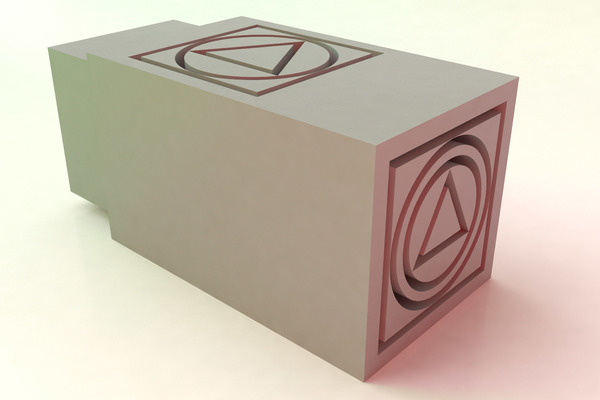}} & \\
\multicolumn{2}{|c|}{\sourcecodeQuoted{Cons:[List (List Bool)]} (2 perspectives)} & \\
\hline
\end{tabu}
}
\caption{Several \mconstructor s.}\label{FIG__Molecular-ADSs__Several-mconstructors}
\end{visupolfigure}

\Cref{FIG__Molecular-ADSs__-Values-with-List-type} shows a number of molecular \ads s built with the \mconstructor s introduced so far. The reader can try to verify that all \mconstructor s indeed fit together, and lead to a well-typed result.

\madawipolA\ is type-safe. \Cref{FIG__-Spatial-Necessity__Weekendday-into-List-Bool} gives a simple example. The power of \madawipolA\, however, lies in how it enforces type-safety of more complex types, such as types that consist of more than one \typeCons, polymorphic types, and recursively defined types. This is the non-trivial part of the design. With the compositional translation of textual types into \typeForm s, as illustrated in the aforementioned, every type can be translated into a \typeForm\ that behaves type-safely, and that reflects the relation between both. We challenge the reader to create a type-incorrect value with the \mconstructor s provided so far. For example, (s)he can try to: join the male joint of \Mcons{Cons:[List (List Bool)]} with a female joint of \Mcons{Cons:[List Bool]} (fails); \Mcons{True} into \Mcons{Cons:[List (List Bool)]} (fails). The reader can also add the \mconstructor s of other instantiations of \sourcecodeQuoted{List a}, such as \sourcecodeQuoted{List (List WeekendDay)}, and then try:  male \Mcons{Cons:[List (List Bool)]} into the side female joint of \Mcons{Cons:[List (List Sat))]} (fails).\coulddo{{| y2018_m08_d25_h12_m00_s31 |} provide a figure with all of these examples.}

Polymorphic types are also type-safe: try fitting \Mcons{Nil} into the side female joint of \emph{any} instantiation of \Mcons{Cons} (always succeeds); \Mcons{Nil} into the top female joint of \Mcons{Cons:[List Bool]} (fails).

\begin{visupolfigure}
\omitForInTestLatexBuilt{
\renewcommand{\scaleFactorA}{0.3}
\renewcommand{\scaleFactorB}{0.45}
\par
\vspace{3mm}
\centering
\tabulinesep=\tabcolsep
\renewcommand{\scaleFactor}{0.090}
\begin{tabu} to \textwidth {|X[1,c,m]X[1,c,m]X[1,c,m]|}
\hline
\href{https://drive.google.com/open?id=1usQ2wWcbksF11PI11dtT1ucQAVxavjUf}{\includegraphics[width=\scaleFactorA\textwidth]{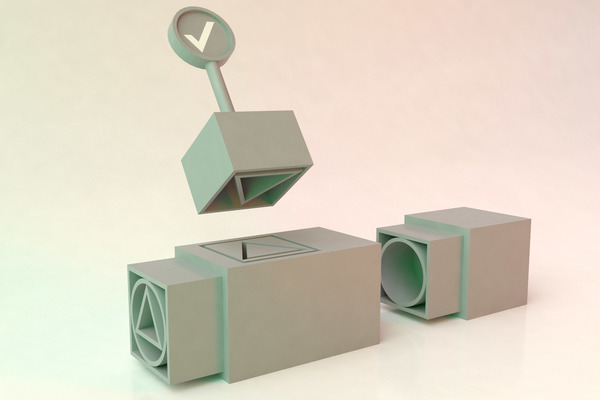}} & %
\href{https://drive.google.com/open?id=1N0ylqhULCTYN-iIu-p_f3nSwhu4qJqOj}{\includegraphics[width=\scaleFactorA\textwidth]{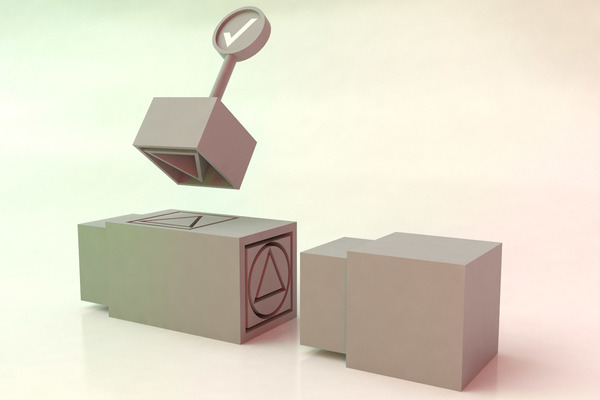}} & %
\href{https://drive.google.com/open?id=1zdy2B5QWaqx3YUFHVtj001fHL9Nmz9Ci}{\includegraphics[width=\scaleFactorA\textwidth]{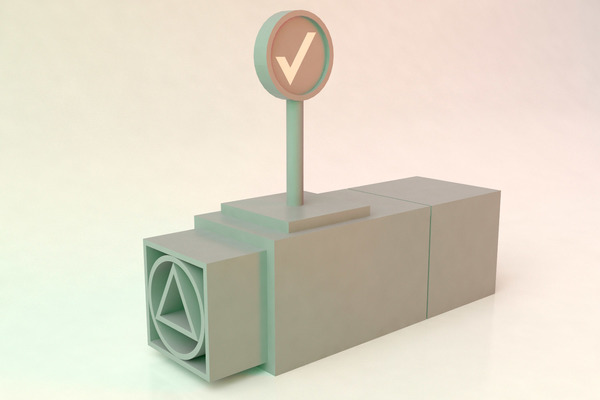}} \\
\multicolumn{3}{|c|}{A maramafied value of type \sourcecodeQuoted{List Bool} (3 perspectives)} \\ 
\hline
\hline
\end{tabu}
\renewcommand{\scaleFactor}{0.14}
\begin{tabu} to \textwidth {|X[1,c,m]X[1,c,m]|}
\href{https://drive.google.com/open?id=1K3q2eGFfoME644GKXiRPxB-1vrm0wNwn}{\includegraphics[width=\scaleFactorB\textwidth]{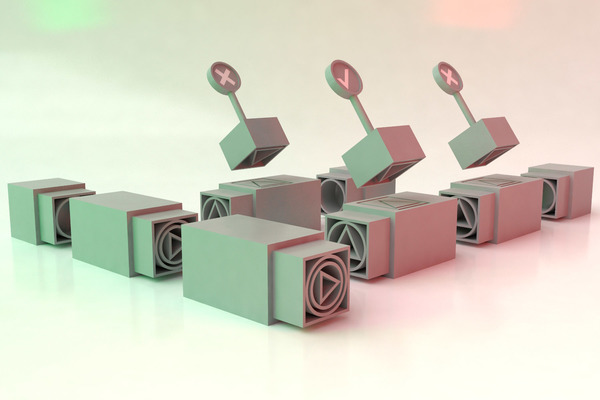}} & %
\href{https://drive.google.com/open?id=1_9dbI-ePhqWdeAx63XiEeAUgRMgQP69a}{\includegraphics[width=\scaleFactorB\textwidth]{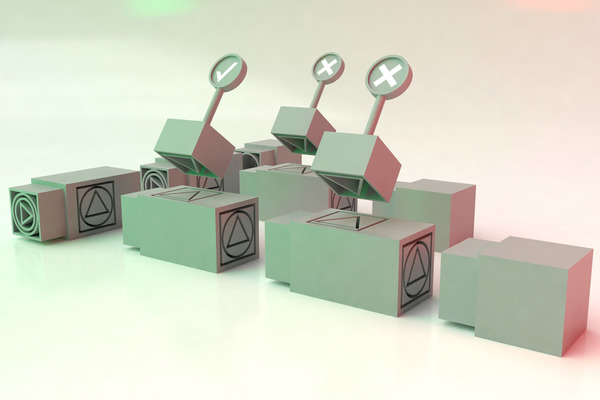}} \\ %
\multicolumn{2}{|c|}{\href{https://drive.google.com/open?id=12TpAt8kzFjYRJ5i-H5B0AiD9K71Mu1Te}{\includegraphics[width=\scaleFactorB\textwidth]{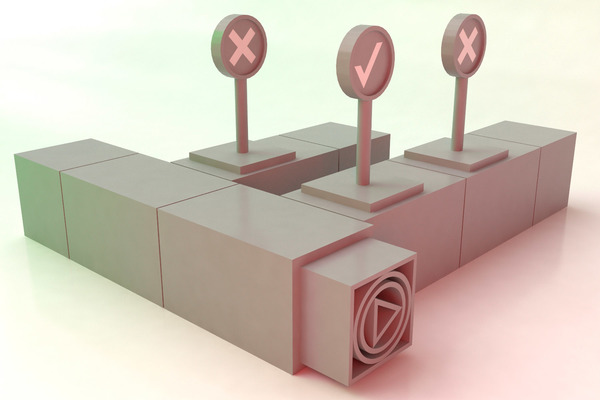}}} \\
\multicolumn{2}{|c|}{A value of type \sourcecodeQuoted{List (List Bool)} (3 perspectives)}  \\ 
\hline
\end{tabu}
}
\caption{Several values with a \sourcecodeQuoted{List}-type.}\label{FIG__Molecular-ADSs__-Values-with-List-type}
\end{visupolfigure}

\begin{visupolfigure}
\omitForInTestLatexBuilt{
\centering
\href{https://drive.google.com/open?id=12UUoU6xTUAmdjXv-Ib-ddij5iywJaRox}{\includegraphics[width=0.8\textwidth]{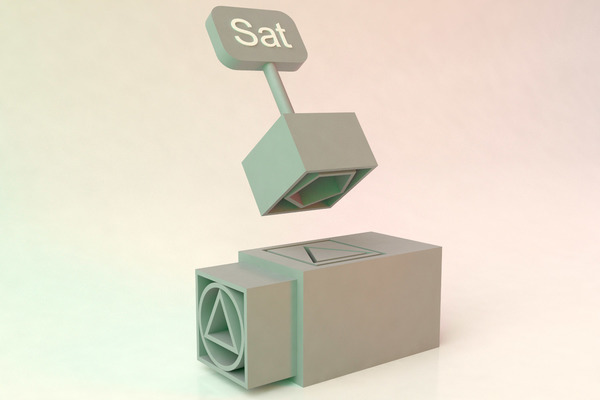}}
}
\caption{Type-safety: \sourcecodeQuoted{Weekendday} does not fit into a list of booleans.}\label{FIG__-Spatial-Necessity__Weekendday-into-List-Bool}
\end{visupolfigure}
\mustdo{{| y2018_m08_d25_h12_m34_s55 |} realise flow of text around figure above.}

\cite{groenouwe2017} contains more examples of \madawipolA.

\section{\PolymorphicSurface s}\label{SEC_--_Polymorphic-surfaces}

\mustdo{how to deal with colours in the article? Perhaps I have to ask Jesse to create a rendering without colours, which still allows one to distinguish the sides.}
A part of \madawipolB\ consists of so called \newterm{\polymorphicSurface s}. These surfaces do not exist as independent objects within \madawipolB, but are, among other things, integrated into polymorphic \mconstructor s. However, for clarity this section explains the \polymorphicSurface s in isolation. The most central part of the explanations focus on the laws that govern the behaviour of the \polymorphicSurface s. The reader is recommended to first try to guess what happens in the figures, and after that read the accompanying explanation.

\Cref{FIG_--ExampsPolySurf--_-Cylinder} shows a gray surface in the center of which a \polymorphicSurface\ $\polySurfConstantFORM$ is attached. (Note that the figure has to be read as a comic strip.) $\polySurfConstantFORM$ has a side with a red colour (in the figure at the top-side) and a side with a blue colour (in the figure at the bottom-side). To demonstrate the behaviour of $\polySurfConstantFORM$, a gray cylinder $C$ is moved into it. As can be seen in the figure, the part of $C$ that touches $\polySurfConstantFORM$ moves downwards with the cylinder. The part that is not touched by $C$, however, moves in the opposite direction over the same distance. $\polySurfConstantFORM$ does not tear in the process, and remains attached to the gray surface. In that sense it behaves as an elastic surface.\coulddo{{| y2018_m05_d10_h18_m26_s53 |} In terms of topology, $\polySurfConstantFORM$ is TODOlookup.}

\begin{visupolfigure}
\omitForInTestLatexBuilt{
\href{https://drive.google.com/open?id=14bxM_zuJdbMekrpymMmqqmvP9tppWX-X}{\includegraphics[width=\columnwidth]{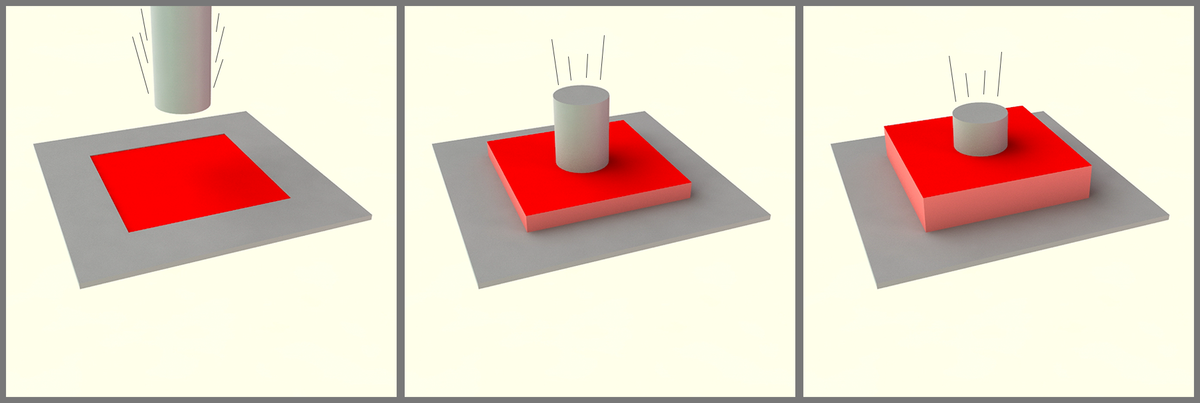}}
\href{https://drive.google.com/open?id=1lwT24wgBb5EI2MJKDwMliIibj9zTY6xq}{\includegraphics[width=\columnwidth]{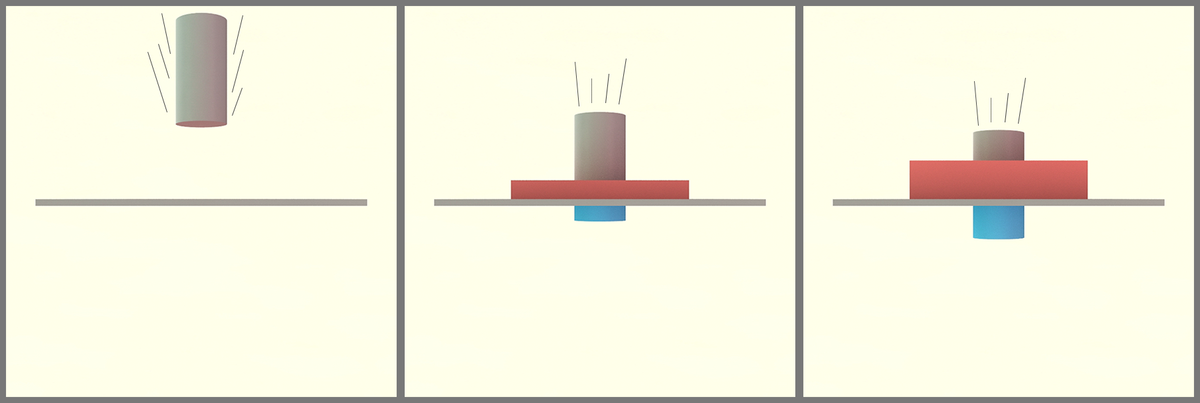}}
\href{https://drive.google.com/open?id=1yiGzj56t2tKrdpFZKx24Mr-PNJARihGJ}{\includegraphics[width=\columnwidth]{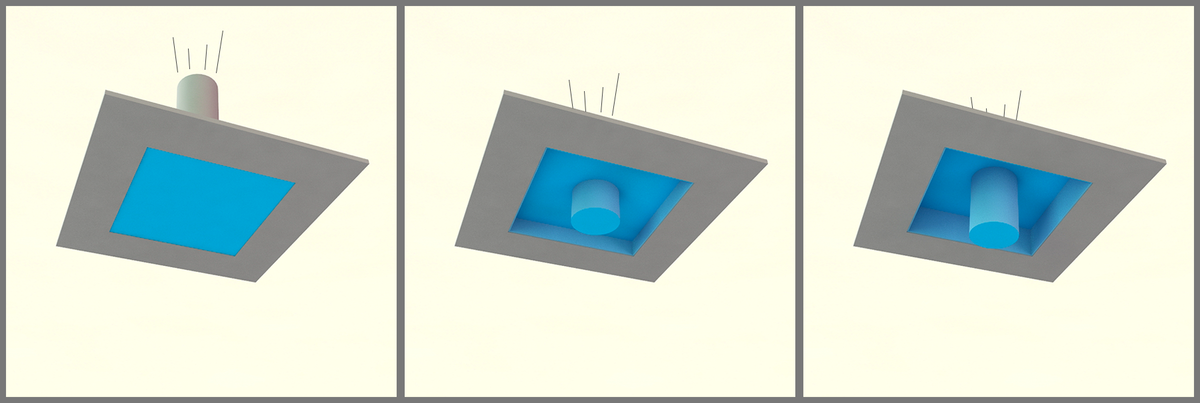}}
}
\caption{A comic strip showing a \polymorphicSurface\ interacting with a cylinder, seen from three perspectives}\label{FIG_--ExampsPolySurf--_-Cylinder}
\end{visupolfigure}

\Cref{FIG_--ExampsPolySurf--_-Mimicking} demonstrates \mimickingPolSurf. A gray surface contains two \polymorphicSurface s: $\polySurfConstantFORM_1$ (left) and $\polySurfConstantFORM_2$ (right). A cylinder $C$ is moved into $\polySurfConstantFORM_1$. $\polySurfConstantFORM_2$ mimics the behaviour of $\polySurfConstantFORM_1$.

\begin{visupolfigure}
\omitForInTestLatexBuilt{
\href{https://drive.google.com/open?id=1TxIEAG8x9DAWH7zUSrl346JPfJzuJUwZ}{\includegraphics[width=\columnwidth]{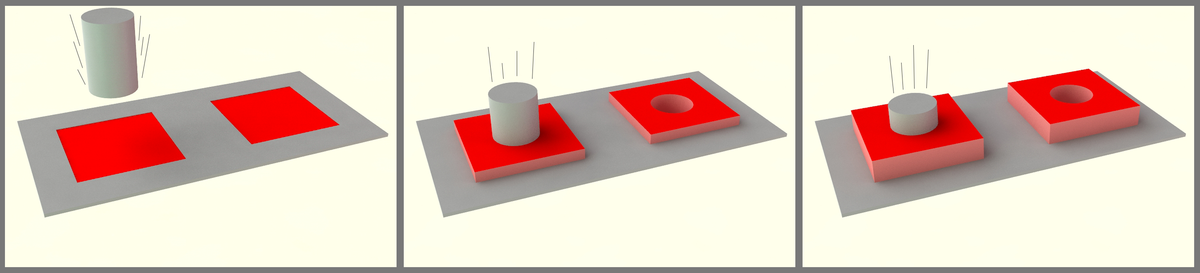}}
\href{https://drive.google.com/open?id=1zzBWif9QanYe4Pk_mLVRzvnmSVpftlUC}{\includegraphics[width=\columnwidth]{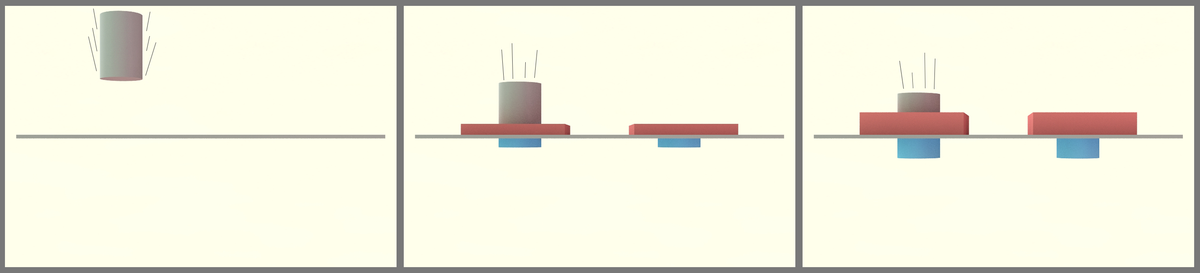}}
\href{https://drive.google.com/open?id=11noEkYk8kPQ7RzK88Mh62RGGf3i2H9kA}{\includegraphics[width=\columnwidth]{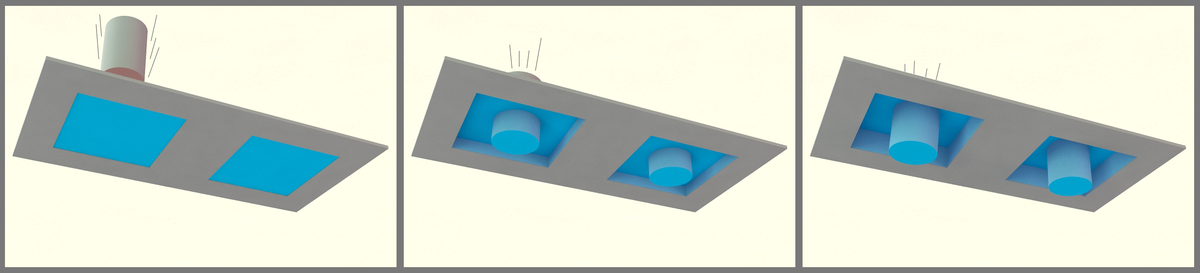}}
}
\caption{Mimicking polymorphic surfaces interacting with a cylinder, seen from three perspectives}\label{FIG_--ExampsPolySurf--_-Mimicking}
\end{visupolfigure}

\Cref{FIG_--ExampsPolySurf--_-TwoPairsMimicking} demonstrates that \mimickingPolSurf\ is confined to \polymorphicSurface s attached to the same object. Each gray surface contains two \polymorphicSurface s. However, only the \polymorphicSurface\ that is on the same gray surface, mimics the behaviour of the \polymorphicSurface\ that receives a cylinder.
\bs{{| y2018_m05_d11_h16_m31_s29 |} at this stage, in this section, do not say anything yet about the law that the part of the polymorphicSurface that is mimicking another surface, is not receptive to receive objects themselves, etc.}
\mustdo{{| y2018_m05_d11_h16_m34_s58 |} think about the exact laws needed for mimicking polymorphic surfaces. This is not so easy. For example, the part of a polymorphic surface that is mimicking another surface is not responsive to objects any more, etc. One things that makes it complicated is that a surface can only be partially in interaction with an object or partially be mimicking. The rest of the surface may then still be receptive to changes.}

\begin{visupolfigure}
\omitForInTestLatexBuilt{
\href{https://drive.google.com/open?id=1cz0LOAQtDk9Kpt64TYmRGpmxYGmrSyiN}{\includegraphics[width=\columnwidth]{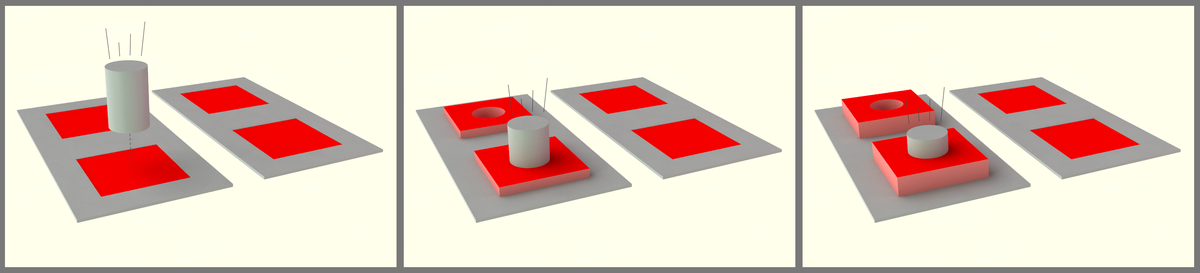}}
\href{https://drive.google.com/open?id=1pG7bi1kim0h138P3uf8aVtyZ6C4pXBFe}{\includegraphics[width=\columnwidth]{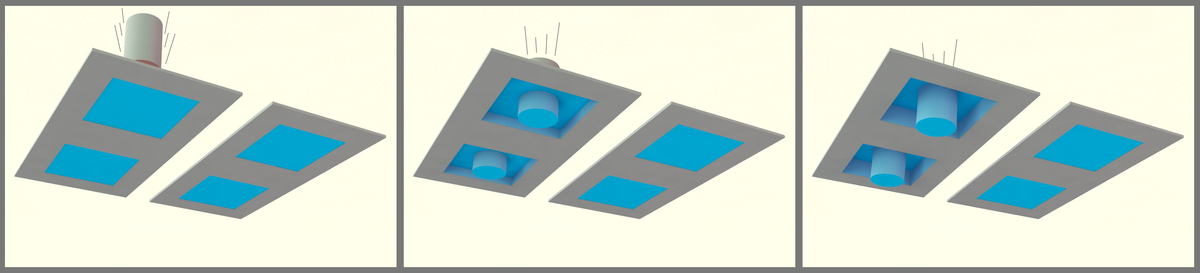}}
}
\caption{Two pairs of polymorphic surfaces, interacting with a cylinder, seen from three perspectives. Each pair is connected to one, separate object.}\label{FIG_--ExampsPolySurf--_-TwoPairsMimicking}
\end{visupolfigure}
\Cref{FIG_--ExampsPolySurf--_-ReversedMimicking} shows two \polymorphicSurface s. $\polySurfConstantFORM_1$ (left) and $\polySurfConstantFORM_2$ (right). $\polySurfConstantFORM_1$'s orientation is the opposite of $\polySurfConstantFORM_2$. One can derive this from the orientation of the colours: in $\polySurfConstantFORM_1$, red is on top, in $\polySurfConstantFORM_2$ it is on the bottom. $\polySurfConstantFORM_2$ mimics the behaviour of $\polySurfConstantFORM_1$, and because of its orientation, the movement is also reversed with respect to the gray surface. Note that this figure is a natural consequence of the laws already exposed in previous figures. It does not introduce any new laws, but is included for clarity.

\begin{visupolfigure}
\omitForInTestLatexBuilt{
\href{https://drive.google.com/open?id=15TpiG4xjv9qb2Ptx3qIRhGnNeMjHnRw5}{\includegraphics[width=\columnwidth]{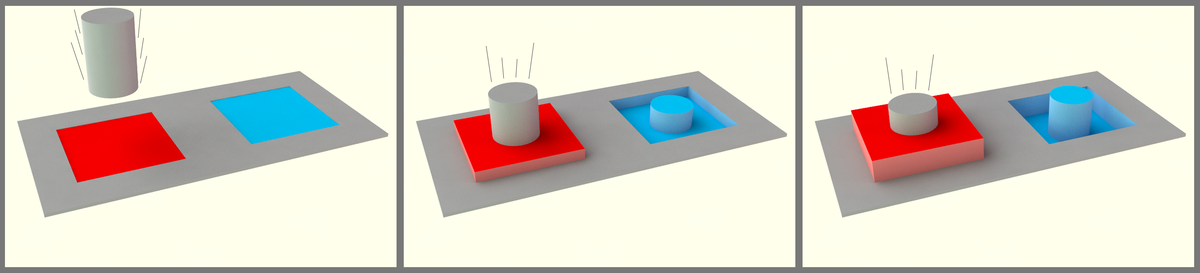}}
\href{https://drive.google.com/open?id=1-UZoQBXKhJYMyY06zKHhwqaIcON9XiMr}{\includegraphics[width=\columnwidth]{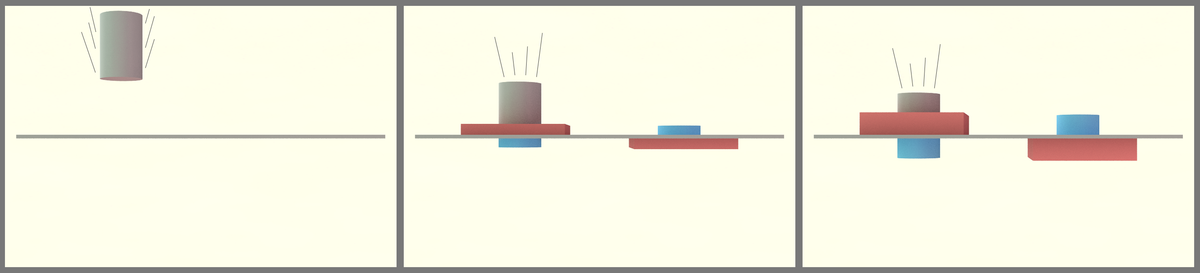}}
\href{https://drive.google.com/open?id=1KEd2RTbzhSnaWLNfFpyN197eCXWC5IpV}{\includegraphics[width=\columnwidth]{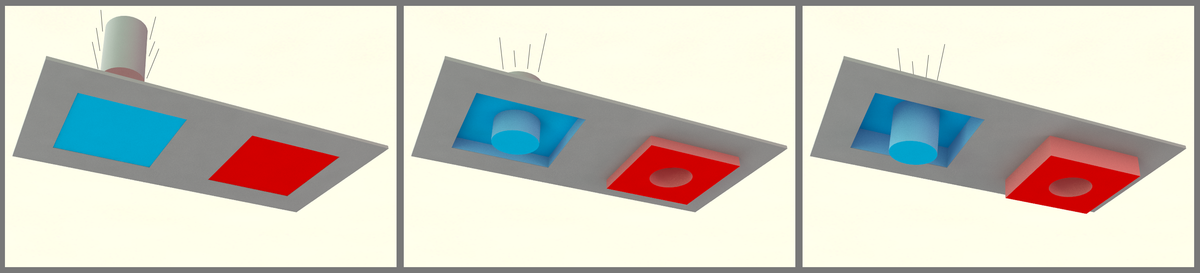}}
}
\caption{Mimicking polymorphic surfaces interacting with a cylinder, seen from three perspectives. The polymorphic surfaces are oriented reversed with respect to each other.}\label{FIG_--ExampsPolySurf--_-ReversedMimicking}
\end{visupolfigure}

\subsection{Differentiation of \polymorphicSurface s}\label{SEC_--_Differentiation}

Some terminology is needed for explanations in the rest of this article. If a \polymorphicSurface\ is fully deformed it is \newterm{\differentiatedCXTpS}. The opposite notion is `\newterm{\undifferentiatedCXTpS}'. A \polymorphicSurface\ always starts in a fully \undifferentiatedCXTpS\ state: a perfectly flat surface as indicated in the examples before. A \polymorphicSurface\ can also be partially differentiated, as will be shown in later examples. The parts of a \polymorphicSurface\ that are differentiated are called \differentiatedRegion s. The opposite notion is `\differentiatedRegion'. A part of a \polymorphicSurface\ that is differentiated as a consequence of mimicking another surface, is called a \mimickingRegion.

If a region of a \polymorphicSurface\ is not mimicking another surface, nor being deformed by an object that is pushed into it then that region is called a \newterm{\freeRegion}. In the examples provided so far, each \polymorphicSurface\ started out as a completely \freeRegion, while no \freeRegion s were left once insertion of an object started. Note that the region of the \polymorphicSurface\ that moves in the opposite direction of the regions being touched by the inserted object, is not a \freeRegion\ either: although it does not touch the inserted object it is clearly being deformed.

\subsection{Law: \mimickingRegion s are rigid}\label{SEC_--_-dif-reg-are-rigid}

A law that is not demonstrated in this article is that \mimickingRegion s of a \polymorphicSurface\ behave like rigid (non-deformable) forms with respect to the objects onto which it exerts a force. As a consequence, if such a \mimickingRegion\ exerts a force onto an \undifferentiatedRegion\ of another \polymorphicSurface, the latter will start to deform as well and become differentiated. Another consequence is that if the other surface is already differentiated, and does not have a matching form, the forms will not fit into each other and movement will halt.\shoulddo{{| y2018_m08_d18_h14_m43_s26 |} add demonstrations, and rewrite the previous sentence.}

\subsection{Law: free regions return to an \undifferentiatedCXTpS\ state}\label{SEC_--_-free_returns_to_undiff}

All examples provided thus far, also work in the reverse direction. When objects are taken out again, the \polymorphicSurface s move back to their original position. These examples can simply be obtained by reading the comic strips in the reverse order. The underlying law is formulated as follows: \freeRegion s return to their \undifferentiatedCXTpS\ state. In other words: any region of a \polymorphicSurface\ that is not being pushed by a shape, nor mimicking another surface, returns to its \undifferentiatedCXTpS\ state.

\subsection{Experimentation}

Note that a quite extensive experimental study has been carried out to test the understandability of the laws mentioned in this section among secondary school students with promising results. \mustdo{TODO include reference to study, and brief summary.}\mustdo{{| y2018_m05_d16_h14_m50_s29 |}}

\section{\madawipolB\ by example}\label{SEC__MadawipolB-by-Example}

As said, \madawipolB\ extends \madawipolA\ with complete parametric polymorphism, i.e.~it also covers polymorphic \mconstructor s with arguments, such as \sourcecodeQuoted{Cons:[List a]}.

For clarity\coreng{verified with google}, in some of the following examples the correspondence between parts of the visual and the textual representation is shown. It is essential to note that these are not put there for definitory purposes: for a user of \madawipolB\ the semantics is contained in the construction possibilities of the building blocks. The correspondence is merely put here to provide the reader of this paper, who is probably well-versed in textual functional languages, a quick insight into\coreng{provide an insight into, verified} the fact that these visualisations indeed exhibit the same relevant behaviours as their textual counterparts.\shoulddo{{| y2018_m05_d09_h18_m44_s42 |} already mentioned this earlier, so abbreviate or omit.}

The essential building block of the extension is formed by \polymorphicSurface s, covered in \cref{SEC_--_Polymorphic-surfaces}.\mustdo{{| y2018_m08_d25_h12_m59_s08 |} provide rest overview.}

In \madawipolB, \polymorphicSurface s are part of polymorphic \mconstructor s. For didactic purposes it easier to first illustrate important aspects of how they function within \mconstructor s in maramafications of the `flexible textual language' \adtPolyOneTooFlexible. It allows `minimal examples'\urlref{https://en.wikipedia.org/wiki/Minimal_Working_Example} to be more minimal.

For the sake of simplicity and the purpose of this article, the \typeConsForm s in this section have been chosen such that 2D cross-sections provide all information needed. This simplifies the creation of examples in static 2D media, such as this article.\done{{| y2018_m08_d25_h13_m03_s01 |} note that 3D however, has a number of advantages, so this does not imply that we could just turn the language into a 2D language.} These cross-sections, however, are not intended as viable alternatives for the 3D models. Among other things, it is harder to discern the different \typeConsForm s in them. For example, the human visual apparatus will immediately recognise a circle that contains a triangle, as two separate objects with a distinct form, while the `ridges and edges' in the 2D cross-sections give it much less to hold on. The users, however, will work with the 3D models.

\mustdo{{| y2018_m08_d31_h12_m44_s55 |} also say something about advantages of the `enscription' design: that an argument of a typeconstructor is visualised as a form *enscribed* by the type constructor: it tallies much more with the intuition that there is a direction of super to sub. An alternative is juxtaposing the forms, which does not have this advantage. Another problem with juxtaposing is scaling, without scaling the joints would become bigger and bigger the more typeconstructors a type contains.}

To achieve the situation that 2D cross-sections contain all information needed, \cref{FIG_--_Several-typeConsForms} redefines the forms of several \typeCons s that have been defined earlier. The latter can be interpreted as viewing a male joint along a line of sight that is perpendicular to the bottom of the joint. We remind the reader that in the electronic version of this document, each image contains clickable links that lead to much higher resolution versions of the image.

\done{{| y2018_m08_d21_h12_m32_s47 |} explain why you now use other forms than in the beginning of the article - e.g. instead of a triangle for bool, the current form. The reason is of course that all info is now also contained in the cross-sections which is easier for this article.}
\begin{visupolfigure}
\omitForInTestLatexBuilt{
\renewcommand{\scaleFactor}{0.30}
\par
\vspace{1cm}
\newcolumntype{C}{>{\centering\arraybackslash} c }  
\centering
\begin{tabular}{CCC}
\href{https://drive.google.com/open?id=1OLRUQc1a7bKUR3zutiFDhgbrhmHo6sfU}{\includegraphics[width=\scaleFactor\textwidth]{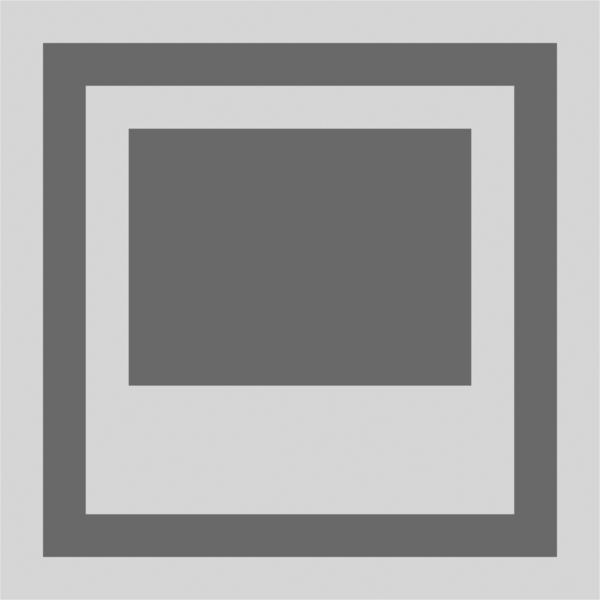}} & %
\href{https://drive.google.com/open?id=1U7G_m6S7eUy_T26qIHLbmwd-9PRq5jSL}{\includegraphics[width=\scaleFactor\textwidth]{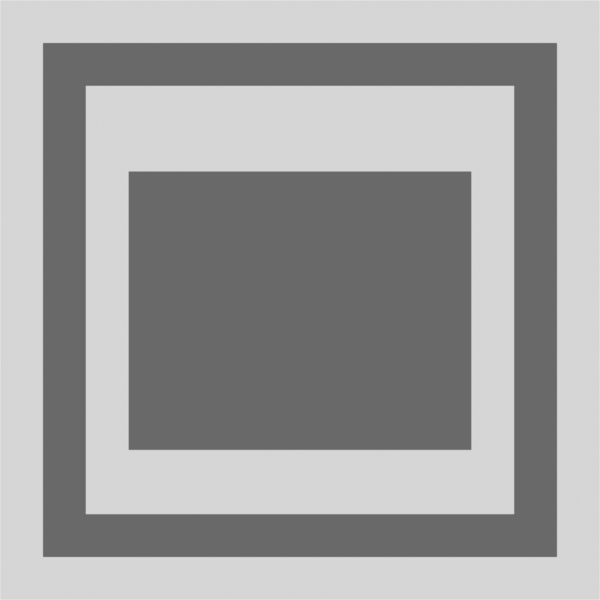}} & %
\href{https://drive.google.com/open?id=1l67XMOveFw0ncfb1jwA-7KFbS82OVAH-}{\includegraphics[width=\scaleFactor\textwidth]{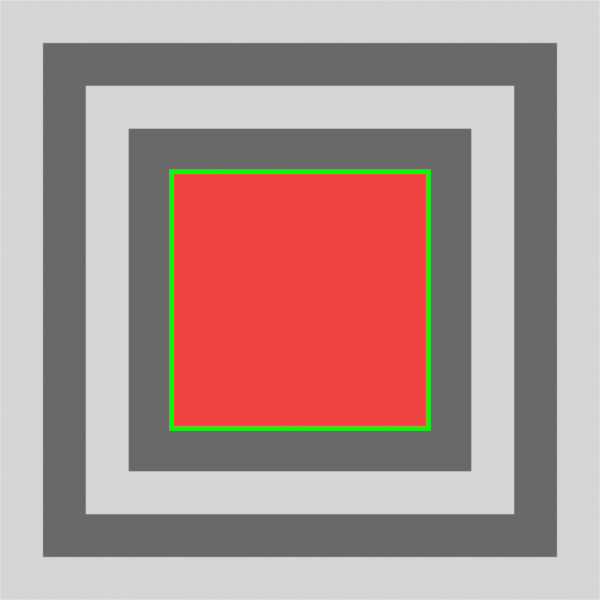}} \\
\sourcecodeQuoted{Colour} & \sourcecodeQuoted{Bool} & \sourcecodeQuoted{List} \\
\end{tabular}
}
\caption{Several \typeConsForm s.}\label{FIG_--_Several-typeConsForms}\shoulddo{{| y2018_m08_d21_h00_m30_s46 |} also add the side cross-section of each joint, and perhaps the joint in 3D.}
\end{visupolfigure}

\Cref{FIG_--_OverviewMconstructorsOne} provides a cross-sectional view on several \mconstructor s with these types. The cross-section is made such that it cuts halfway through each joint. Note the red and the blue lines indicating the orientation of each \polymorphicSurface.\commentByDouweSchulte{{| y2018_m08_d31 |} lines are purple for him, contrast problem?} These lines, in reality have no thickness as they are part of the same surface. As a compromise, these parallel lines are drawn such that the center of the combined lines aligns with the real line. Specifically in later figures, where \mconstructor s are joined, and up to four lines stack up, this is a workable compromise.

\shoulddo{{| y2018_m05_d13_h16_m38_s36 |} say somewhere, however, that 2D does not suffice for the final design and why.} In the remainder of this article, the prefix `M-' indicates `maramafied'. For example, the \mconstructor\ (maramafied constructor) of \sourcecodeQuoted{FlexiCons} from \cref{EXAMPLE_--_AlgepolyOnePlus} is indicated with M-\sourcecodeQuoted{FlexiCons}.

Before starting with the elucidating examples, \cref{FIG_--_OverviewMconstructorsOne} provides an overview of all \mconstructor s used in the examples, for later reference.

\begin{visupolfigure}
\omitForInTestLatexBuilt{
\par
\vspace{1cm}
\newcolumntype{C}{>{\centering\arraybackslash} c }  
\centering
\renewcommand{\scaleFactorA}{0.22}
\renewcommand{\scaleFactorB}{0.48}
\newcommand{\figRed}{\href{https://drive.google.com/open?id=1oRpyLCpehG6z5gJQIXjnJoQL7LV-Ygii}{\includegraphics[width=\scaleFactorA\textwidth]{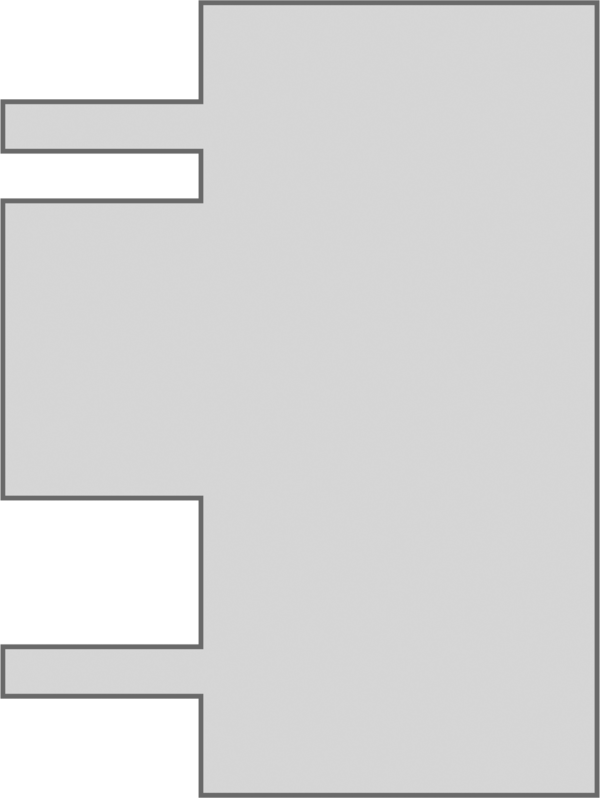}}}
\newcommand{\figSimpleFemCons}{\href{https://drive.google.com/open?id=1BqzQPnePaj-6X3NLWUR5k_xAoaDYNYCa}{\includegraphics[width=\scaleFactorA\textwidth]{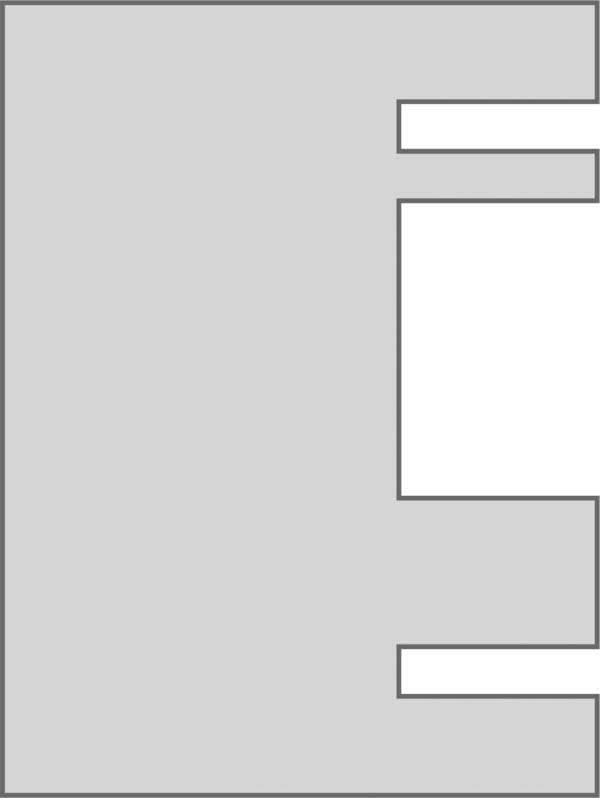}}}
\newcommand{\figFlexiCons}{\href{https://drive.google.com/open?id=1BmCgvMvewa9Akp9afpABQaf1ZG-5QuuZ}{\includegraphics[width=\scaleFactorA\textwidth]{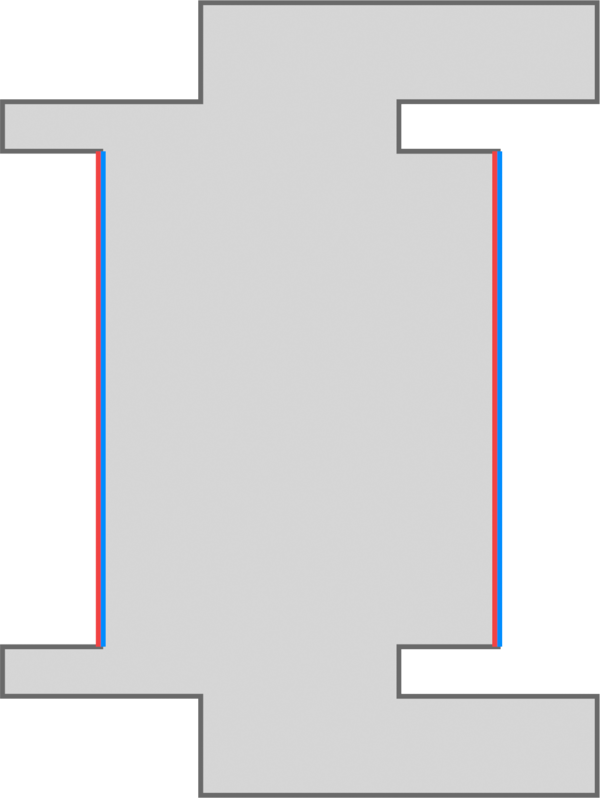}}}
\newcommand{\figPolyCons}{\href{https://drive.google.com/open?id=1QWp9CCCeEQAbXa3QOP3qb1oSAqxxf3Y7}{\includegraphics[width=\scaleFactorA\textwidth]{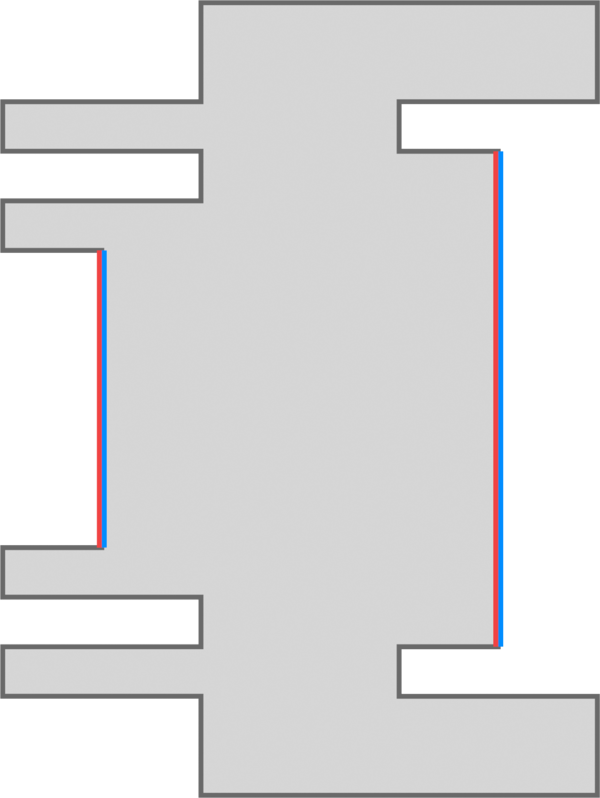}}}
\newcommand{\figTrue}{\href{https://drive.google.com/open?id=1vYNs7Nsi-cdh3S9KopCQr4U_zgI4t0PQ}{\includegraphics[width=\scaleFactorA\textwidth]{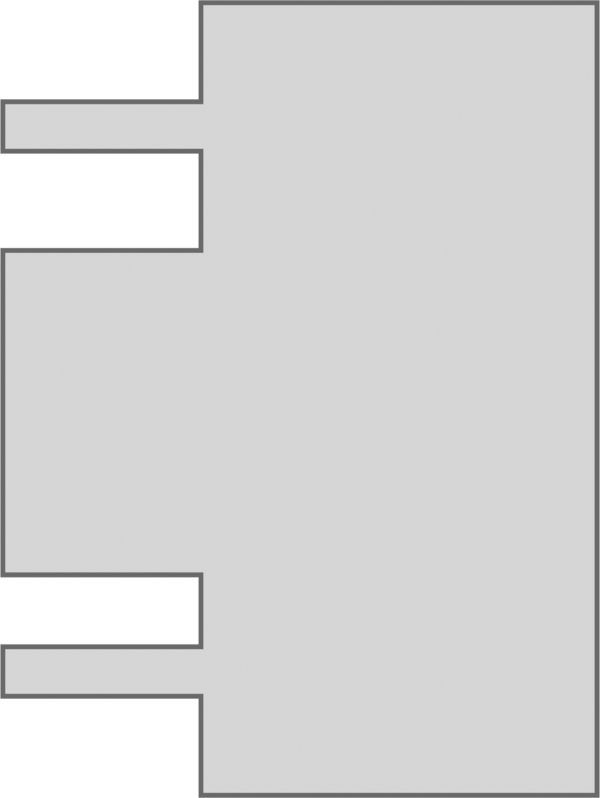}}}
\begin{tabular}{CCCC}
\figRed & %
\figSimpleFemCons & %
\figFlexiCons & %
\figPolyCons \\
\Mcons{Red} & \Mcons{SimpleFemCons} & \Mcons{FlexiCons} & \Mcons{PolyCons} \\
\sourcecodeQuoted{Colour <-} & \sourcecodeQuoted{<- Colour} & \sourcecodeQuoted{a <- a} & \sourcecodeQuoted{SimpleType a <- a} \\
\figTrue & %
\multicolumn{2}{c}{\href{https://drive.google.com/open?id=1zRkvhtMlYzZ3bGtdNqwQNv7aI-CbyJK4}{\includegraphics[width=\scaleFactorB\textwidth]{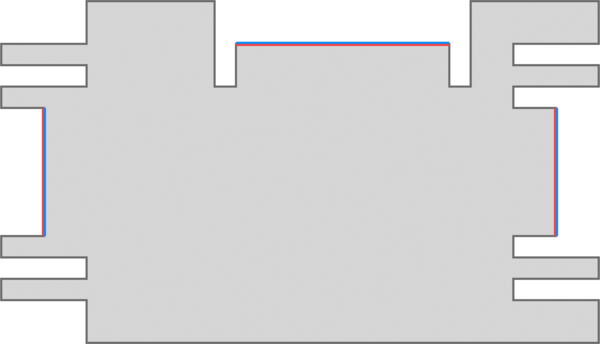}}} &
\href{https://drive.google.com/open?id=1AIBga8e284BU59rY8Yb_spJCzntME7vd}{\includegraphics[width=\scaleFactorA\textwidth]{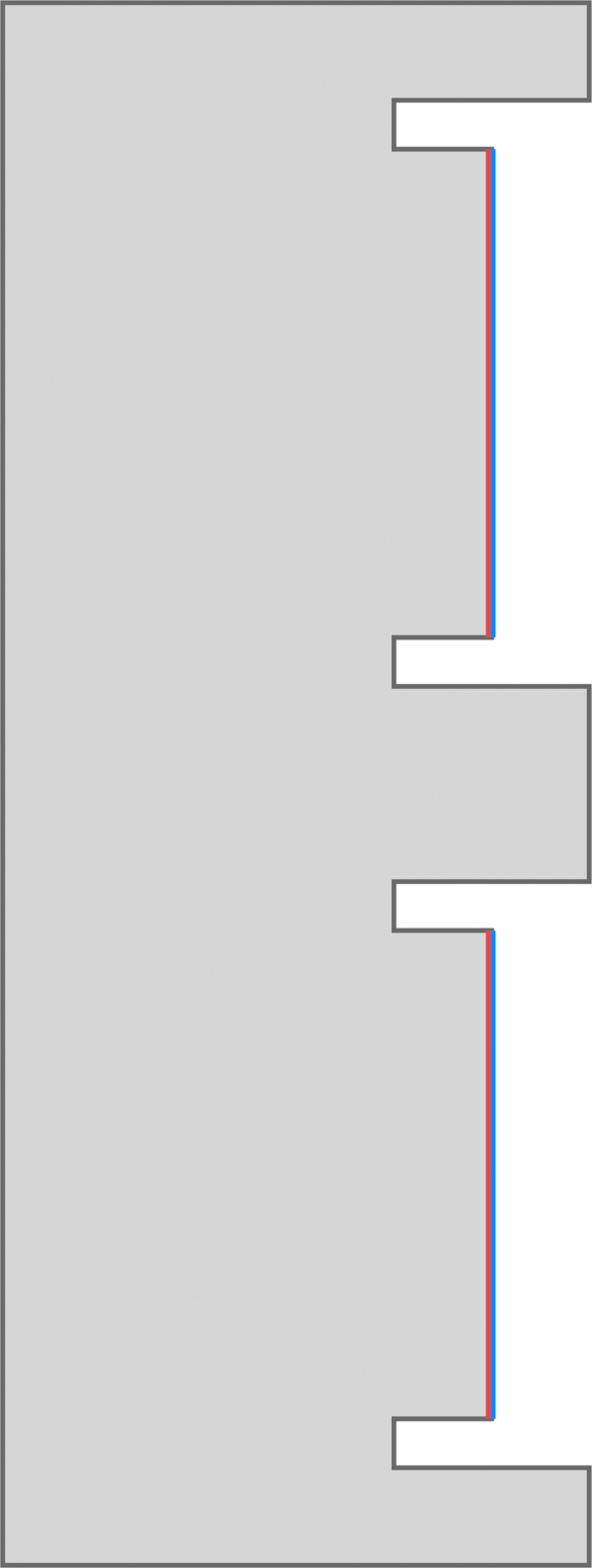}}\\
\Mcons{True} & \Mcons{Cons} & \multicolumn{2}{c}{\Mcons{SimplePairCons}}\\
\sourcecodeQuoted{Bool <-} & \sourcecodeQuoted{List a <- a (List a)} & \multicolumn{2}{c}{\sourcecodeQuoted{<- a a}}\\
\end{tabular}
}
\caption{Overview of \mconstructor s. Their name and type are written beneath them.}\label{FIG_--_OverviewMconstructorsOne}
\end{visupolfigure}
\mustdo{{| y2018_m08_d25_h15_m15_s20 |} brief explanation coloured surfaces.}
\shoulddo{{| y2018_m08_d17_h21_m30_s13 |} provide examples that demonstrate how a textual type corresponds with a jointForm, how they can be constructed.
Show:
simple types including parameterised:
Bool
Colour
a

Type constructor application
SimpleType a
SimpleType Bool
SimpleType SimpleType Red

Make clear that repetitive.

Note that repeated \typeCons\ applications can occur hard-coded in Clean etc. as well, for example:
::SpecialList a = SpecialCons a (SpecialList SpecialList a) | Nil
}

\newcommand{\introducesNewLaw}{introduces a new law}
\newcommand{\IntroducesNewLaw}{Introduces a new law}
\newcommand{\demoExistingLaws}{illustrates existing laws}
\newcommand{\DemoExistingLaws}{Illustrates existing laws}
\newcommand{\implicationExistingLaws}{implications of existing laws}
\newcommand{\ImplicationExistingLaws}{Implications of existing laws}
The rest of this section introduces the behaviours of these \mconstructor s by providing examples. It will indicate for each example whether it ``\introducesNewLaw'', or whether it merely ``\demoExistingLaws''. The latter type of examples should be perfectly predictable and are, among other things, added to allow the reader to verify his or her understanding.

\subsection{Reverse reading of examples}

Let the reader be reminded that all examples provided in the following sections also work in the reverse direction, following the law that has been explained in \cref{SEC_--_-free_returns_to_undiff}.

\subsection{Law: mimicking \polymorphicSurface s (intra-\mconstructor\ type-propagation)}
\mustdo{{| y2018_m08_d25_h15_m15_s48 |} brief explanation of the way the lines are represented when constructors are joined (e.g. red-blue-red-blue). and the compromise that all lines/surfaces are in fact infinitely thin.}

(\ImplicationExistingLaws.) \Cref{FIG_--_-Intra-type-propagation} demonstrates type propagation within a \mconstructor, i.e.~intra \mconstructor\ type-propagation. Simply by following the laws of \polymorphicSurface s from \cref{SEC_--_Polymorphic-surfaces}, and the chosen forms of the joints, the type propagation takes place as expected in these examples. It first shows an \href{https://drive.google.com/open?id=1BmCgvMvewa9Akp9afpABQaf1ZG-5QuuZ}{\Mcons{FlexiCons}} (left) and an \href{https://drive.google.com/open?id=1oRpyLCpehG6z5gJQIXjnJoQL7LV-Ygii}{\Mcons{Red}} (right). The corresponding textual versions have been defined in \cref{EXAMPLE_--_AlgepolyOnePlus}. The fixed male joint of \href{https://drive.google.com/open?id=1oRpyLCpehG6z5gJQIXjnJoQL7LV-Ygii}{\Mcons{Red}} moves into the (polymorphic) female joint of \href{https://drive.google.com/open?id=1BmCgvMvewa9Akp9afpABQaf1ZG-5QuuZ}{\Mcons{FlexiCons}}. As expected, the \polymorphicSurface\ in the male joint of \href{https://drive.google.com/open?id=1BmCgvMvewa9Akp9afpABQaf1ZG-5QuuZ}{\Mcons{FlexiCons}} mimics the \polymorphicSurface\ that is being manipulated: the one in the female joint. As a consequence of the inverted orientation of the \polymorphicSurface\ of the female joint with respect to the male joint, the form of the male joint will become the exact inverse of the female joint when it mimics the latter. The male joint of \href{https://drive.google.com/open?id=1BmCgvMvewa9Akp9afpABQaf1ZG-5QuuZ}{\Mcons{FlexiCons}} has now taken the exact shape of the male joint of \href{https://drive.google.com/open?id=1oRpyLCpehG6z5gJQIXjnJoQL7LV-Ygii}{\Mcons{Red}}. This is perfectly consistent with the typing information expressed in the textual versions of the constructors, and is a first step in illustrating how type-information propagates through \mconstructor s. Note that this also partially explains why an undifferentiated\done{{| y2018_m05_d16_h20_m00_s44 |} define undifferentiated.} \polymorphicSurface s is positioned halfway the joint, and why the \polymorphicSurface\ moves in two directions at the same time.\shoulddo{ This will be explained in more detail in \cref{SEC__Definition}.\mustdo{{| y2018_m05_d15_h18_m12_s11 |} refine reference in last sentence as soon as this has been written.}}
\begin{visupolfigure}
\omitForInTestLatexBuilt{
\shoulddo{{| y2018_m12_d10_h18_m47_s17 |} better to base width of scaleFactorB on scaleFactor.}




\renewcommand{\scaleFactor}{0.45}
\renewcommand{\scaleFactorB}{0.35}
\vspace{3mm}
\centering
\tabulinesep=\tabcolsep
\begin{tabu} to \textwidth {|X[1,c,m]|X[1,c,m]|}
\hline
\href{https://drive.google.com/open?id=16_iQXzLqzfbwK-4DBv99E6U7l6dt5DTX}{\includegraphics[width=\scaleFactor\textwidth]{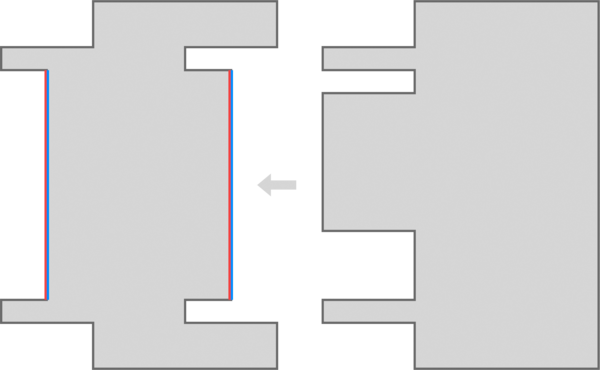}} & %
\href{https://drive.google.com/open?id=19b7DFnPm6alxADcpYGRL9S7iI9f4I65C}{\includegraphics[width=\scaleFactorB\textwidth]{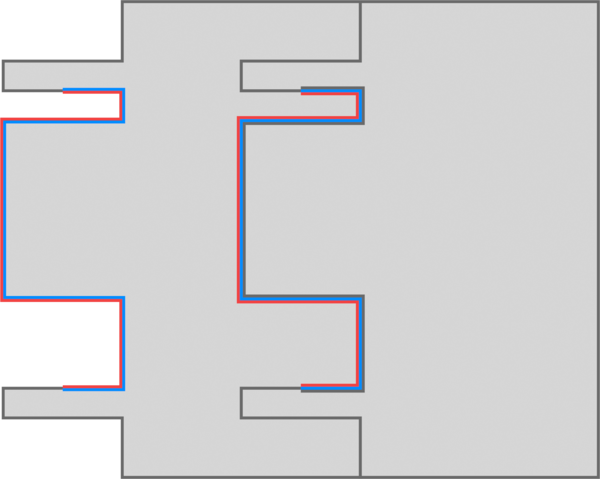}} \\
\hline
\multicolumn{2}{|c|}{\Mcons{FlexiCons:[a<-a]} receives \Mcons{Red:[Colour<-]}.}\\
\hline
\href{https://drive.google.com/open?id=1wRM0vTOQ_w3NanSEi7Jx-sYEiDvrxWxK}{\includegraphics[width=\scaleFactor\textwidth]{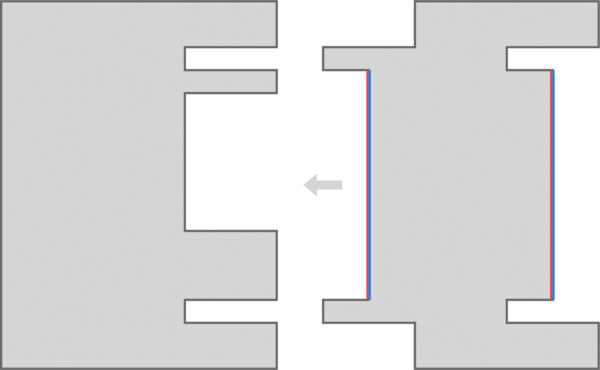}} & %
\href{https://drive.google.com/open?id=1oAGwndffsgRXy-NKQmMW65LyOMDylhNB}{\includegraphics[width=\scaleFactorB\textwidth]{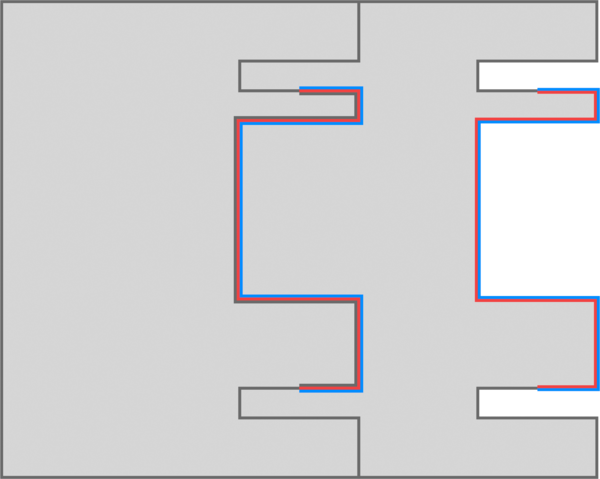}} \\
\hline
\multicolumn{2}{|c|}{\Mcons{SimpleFemCons:[<-Colour]} receives \Mcons{FlexiCons:[a<-a])}}\\
\hline
\href{https://drive.google.com/open?id=1fJhv-ZbyWLwVK4VeZ1_dn-_ntpn2PMZD}{\includegraphics[width=\scaleFactor\textwidth]{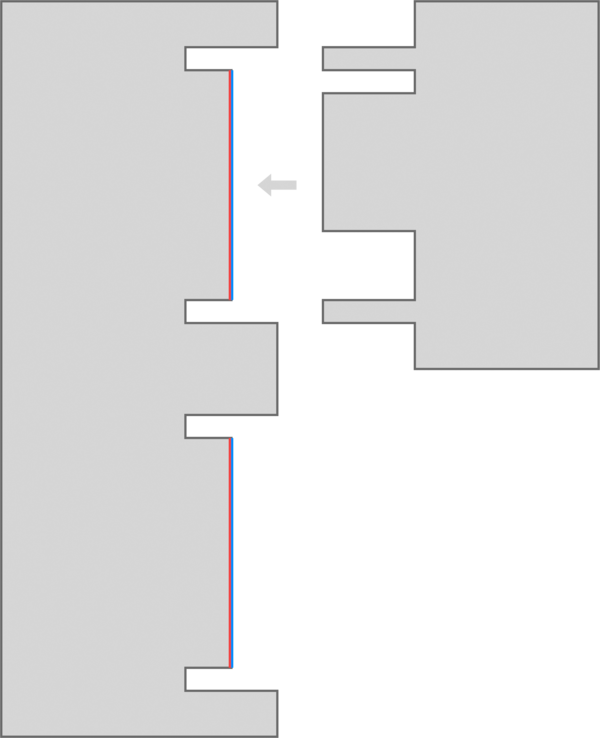}} & %
\href{https://drive.google.com/open?id=1BjEuU9Xuwrd3XVOXnPGbzgyuD8ruaE-2}{\includegraphics[width=\scaleFactorB\textwidth]{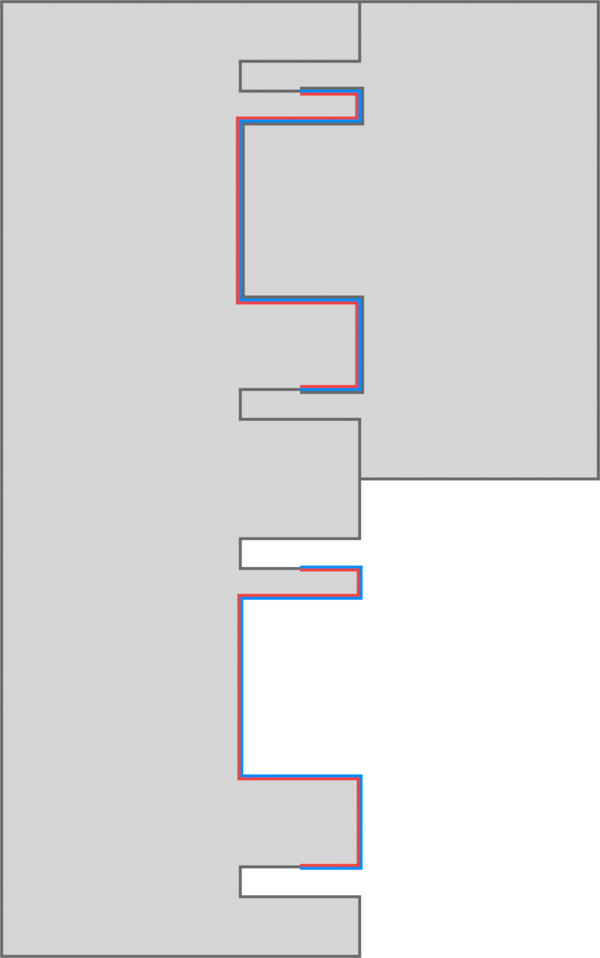}} \\
\hline
\multicolumn{2}{|c|}{\Mcons{SimplePairCons:[<-a a]} receives \Mcons{Red:[Colour<-]}}\\
\hline
\end{tabu}
}
\caption{Comic-strips demonstrating intra-\mconstructor\ type-propagation.}\label{FIG_--_-Intra-type-propagation}\mustdo{{| y2018_m08_d25_h15_m22_s07 |} rename label.}
\end{visupolfigure}

The rest of \cref{FIG_--_-Intra-type-propagation} illustrates that the intra-\mconstructor\ propagation takes place as expected, whether it is propagation from a male joint to a female joint, from a female joint to a male joint, or from a female joint to a female joint.

\shoulddo{{| y2018_m05_d15_h19_m12_s29 |} nice to add an additional example in which one mconstructor has two female joints and one male joint, and where something is being put into one female joint. This is to illustrate that the male and the female joint mimic in a reversed way. Then also state that any combination works correctly, and that this is left as an exercise for the reader.}

\subsection{Inter-\mconstructor\ type propagation}
\shoulddo{{| y2018_m08_d18_h12_m11_s39 |} I was inclined to call it mechanical interaction, however, this can be confusing because the polymoprhic surfaces do not exactly follow the laws of mechanical interaction (because of the opposing movement.}

(\DemoExistingLaws.) \Cref{FIG_--_FlexiCons-into-Two-FlexiConses} shows how type-information propagates through several \mconstructor s: i.e.~inter-\mconstructor\ type propagation. \href{https://drive.google.com/open?id=1oRpyLCpehG6z5gJQIXjnJoQL7LV-Ygii}{\Mcons{Red}} moves into the right \href{https://drive.google.com/open?id=1BmCgvMvewa9Akp9afpABQaf1ZG-5QuuZ}{\Mcons{FlexiCons}} $\MflexCFORM_1$. The \polymorphicSurface\ at the left of $\MflexCFORM_1$, mimics the other one of $\MflexCFORM_1$. The right \polymorphicSurface\ of $\MflexCFORM_2$ (the right $\MflexCFORM_2$) touches the changing \polymorphicSurface, and responds by following the laws that were introduced in \cref{SEC_--_Polymorphic-surfaces}. Therefore, it gets exactly the same shape as the changing \polymorphicSurface. The left \polymorphicSurface\ of $\MflexCFORM_2$ mimics the right \polymorphicSurface\ of $\MflexCFORM_2$. This results in a final situation that is, type-wise, in perfect agreement with the textual counterparts of the \mconstructor s.
The figure also demonstrates that \mimickingRegion s of a \polymorphicSurface\ act as rigid objects (also see \cref{SEC_--_-dif-reg-are-rigid}).
\mustdo{{| y2018_m05_d15_h18_m53_s34 |} add labels of joints, surfaces and constructors etc. as an overlay to the figures. That will make the texts much more readable I think.}

\begin{visupolfigure}
\omitForInTestLatexBuilt{
\renewcommand{\scaleFactorA}{0.95}
\renewcommand{\scaleFactorB}{0.8}
\vspace{3mm}
\centering
\tabulinesep=\tabcolsep
\begin{tabu} to \textwidth {|X[1,c,m]|}
\hline
\href{https://drive.google.com/open?id=1gwNbpWcB5OgDDuK0QVAEWq9iaBrxPmW_}{\includegraphics[width=\scaleFactorA\textwidth]{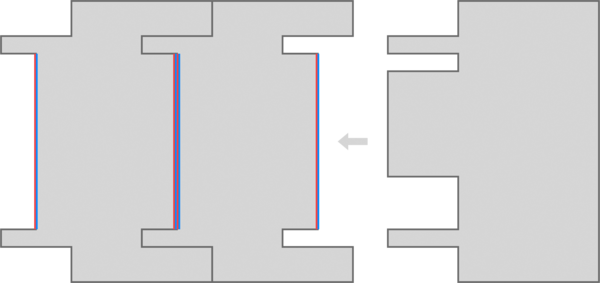}} \\ %
\href{https://drive.google.com/open?id=1rxcTo1PLMAlLmS8FEoTq8fMsIgfZx1wu}{\includegraphics[width=\scaleFactorB\textwidth]{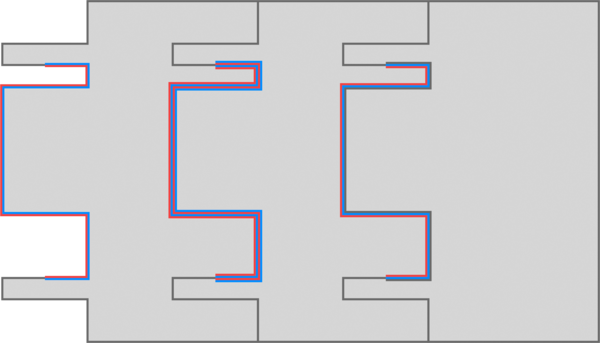}} \\
\hline
\end{tabu}
}
\caption{Two already joined \Mcons{FlexiCons}es receive an \Mcons{Red}. The result's textual equivalent is \sourcecodeQuoted{FlexiCons (FlexiCons Red)}.}\label{FIG_--_FlexiCons-into-Two-FlexiConses}
\end{visupolfigure}

\subsection{Law: scaling of \polymorphicSurface s}
\shoulddo{{| y2018_m08_d25_h16_m48_s52 |} in a future version this should be part of explaining the laws of the polymorphicSurfaces in isolation.}

(\IntroducesNewLaw.) \Cref{FIG_--_Red-into-PolyCons} demonstrates scaling of \polymorphicSurface s. The \polymorphicSurface\ at the left ($\polySurfConstantFORM_2$) is smaller than the one at the right ($\polySurfConstantFORM_1$). As expected, $\polySurfConstantFORM_2$ mimics $\polySurfConstantFORM_1$, however its vertical size is scaled down. The scaling factor is, in this case, equal to the ratio between the sizes of both \polymorphicSurface s. This scaling process, a linear transformation to be precise, plays the central role in the maramafication of the application of type constructors (i.e.~creating joints of types such as \sourcecodeQuoted{(SimpleType a)} or \sourcecodeQuoted{(SimpleType (List (SimpleType a)))}. For example, the joint that corresponds with \sourcecodeQuoted{(SimpleType a)} consists of an alignment square (the two outer `pins'), the \typeCons\ \sourcecodeQuoted{SimpleType} (the inner two `pins'), and the type parameter \sourcecodeQuoted{a} (the scaled down \polymorphicSurface). This will be explained in more detail in \cref{SEC__Definition-MadawipolB__typeConsForms,SEC__Definition-MadawipolB__typeForms}.\mustdo{{| y2018_m05_d16_h14_m13_s10 |}}\shoulddo{{| y2018_m05_d16_h14_m06_s33 |} I try to explain too much in one picture. Create other examples to demonstrate how \typeCons\ application works in marama.}

\mustdo{{| y2018_m05_d16_h12_m56_s50 |}}
\begin{visupolfigure}
\omitForInTestLatexBuilt{
\renewcommand{\scaleFactorA}{0.95}
\renewcommand{\scaleFactorB}{0.8}
\vspace{3mm}
\centering
\tabulinesep=\tabcolsep
\begin{tabu} to \textwidth {|X[1,c,m]|}
\hline
\href{https://drive.google.com/open?id=1CcHAPf0Rjx4X89kzxKNCxZ1hC5NRLrrA}{\includegraphics[width=\scaleFactorA\textwidth]{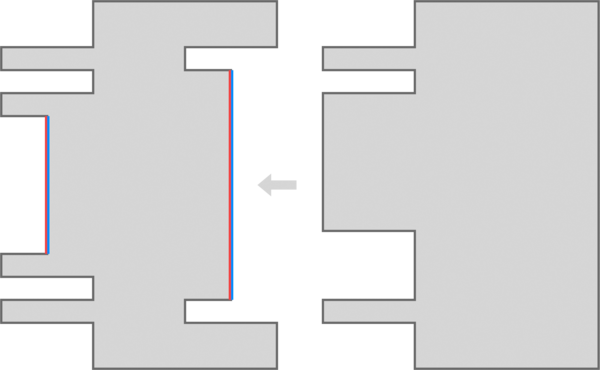}} \\ %
\href{https://drive.google.com/open?id=1PQC_2-r_eLeT8d_JzmVTmjk5yiHTeFkr}{\includegraphics[width=\scaleFactorB\textwidth]{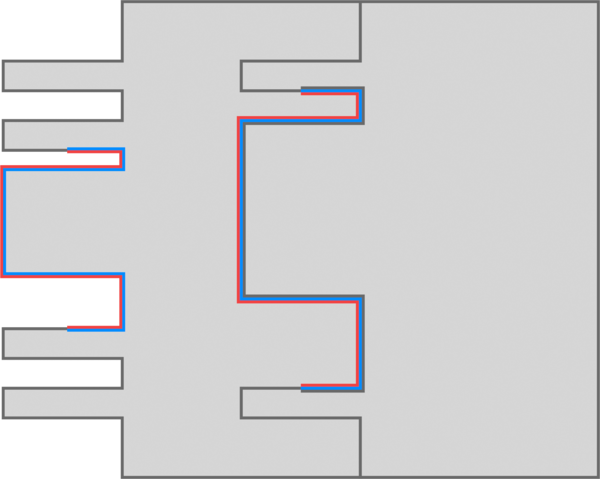}} \\
\hline
\end{tabu}
}
\caption{\sourcecodeQuoted{PolyCons Red}.}\label{FIG_--_Red-into-PolyCons}
\end{visupolfigure}

\mustdo{{| y2018_m08_d18_h11_m19_s43 |} next paragraph seems double!?}

\subsection{Joining `super'- and `sub'types}

\mustdo{{| y2018_m08_d30_h19_m57_s42 |} what is the correct name for super and subtypes in FP, e.g. what is the relation between List a and List?}
\done{{| y2018_m08_d29_h20_m07_s25 |} rename this back to joining super and subtypes. The current title (unifiable types are joinable) should be treated in the proof section.}
\mustdo{{| y2018_m08_d29_h20_m04_s42 |} further check and rewerite this section, also add examples using the typeconsform and typeform figures. But perhaps these can better be placed in the section with the proof strategy of type-safety of madawipolB.}
(\IntroducesNewLaw.) \Cref{FIG_--_M-PolyCons_M-PolyCons} shows an example of joining \polymorphicSurface s where the female joint is a strict super-type of the male joint\mustdo{is this correct terminology, strict supertype in this context?}. In other words, the types are different, yet unifiable. In the figure, a female joint with type \sourcecodeQuoted{a} is connected with a male joint with type \sourcecodeQuoted{SimpleType a}. As one can see, the undifferentiated region of a polymorphic surface that meets an undifferentiated region of the polymorphic surface of the opposing joint, remains in an undifferentiated state. The rest of the polymorphic surface takes the shape of the opposing joint.\done{{| y2018_m08_d17_h21_m19_s15 |} introduce the notion undifferentiated in earlier examples.}\shoulddo{{| y2018_m08_d17_h21_m21_s11 |} perhaps explicitly define all the laws, using a laws environment in latex.}

(\ImplicationExistingLaws.) It also demonstrates that the \mconstructor s can fully deal with recursively defined types. (These are types in which a \typeCons\ occurs both in the left-hand side and the right-hand side of the \ADTD, such as in the definition of \sourcecodeQuoted{PolyCons}.) The left-most joint takes the shape that corresponds with the type \sourcecodeQuoted{SimpleType (SimpleType a)}. The reader is invited to verify that if yet another \Mcons{PolyCons} would be inserted in the right \Mcons{PolyCons}, it would lead to the left-most joint taking the form \Mtype{SimpleType (SimpleType (SimpleType a))}. The notation \Mtype{Type} means the \typeForm\ of the given type.\mustdo{{| y2018_m09_d04_h19_m03_s06 |} always check whether this notiation intro is still in a good place, move it to the first time you use it. Do not remove this mustdo.}

\Cref{FIG_--_M-PolyCons_M-PolyCons} introduces an additional law.

\begin{visupolfigure}
\omitForInTestLatexBuilt{
\renewcommand{\scaleFactorA}{0.47}
\renewcommand{\scaleFactorB}{0.38}
\vspace{3mm}
\centering
\tabulinesep=\tabcolsep
\begin{tabu} to \textwidth {|X[1,c,m]|X[1,c,m]|}
\hline
\href{https://drive.google.com/open?id=18xkoShhOO05qJ_B0hPOrV5QA6Ejs_V-3}{\includegraphics[width=\scaleFactorA\textwidth]{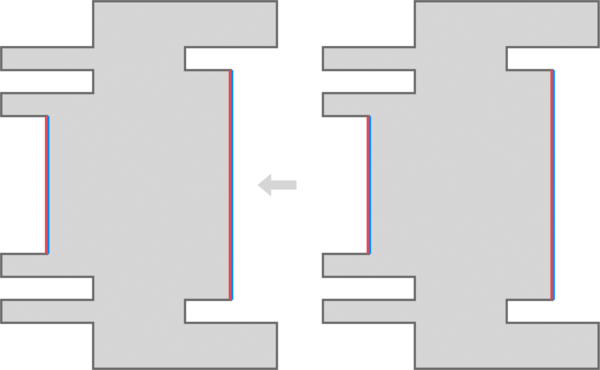}} & %
\href{https://drive.google.com/open?id=1PtJcBQbvYp8JcFmdBqx6JktveEuTO0H8}{\includegraphics[width=\scaleFactorB\textwidth]{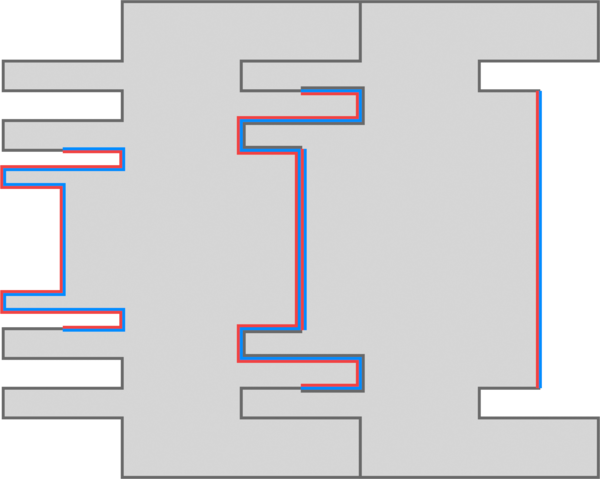}} \\
\hline
\multicolumn{2}{|c|}{\sourcecodeQuoted{PolyCons (PolyCons \_)}}\\
\hline
\end{tabu}
}
\caption{\sourcecodeQuoted{PolyCons (PolyCons \_)}}\label{FIG_--_M-PolyCons_M-PolyCons}
\end{visupolfigure}

(\DemoExistingLaws.) \Cref{FIG_--_M-PolyCons_PolyCons_Red} shows a more complex example of scaling during type-propagation. By inserting \href{https://drive.google.com/open?id=1oRpyLCpehG6z5gJQIXjnJoQL7LV-Ygii}{\Mcons{Red}} into the righter \Mcons{PolyCons} of \cref{FIG_--_M-PolyCons_PolyCons_Red}, scaling takes place two times. This results in the left-most joint to take on the shape that corresponds to the type \sourcecodeQuoted{SimpleType (SimpleType Red)}.

\begin{visupolfigure}
\omitForInTestLatexBuilt{
\renewcommand{\localcaption}{\sourcecodeQuoted{PolyCons (PolyCons Red)}}
\renewcommand{\figA}{\href{https://drive.google.com/open?id=1k5u2kPsY7qMxz0RjOgUk2gP5saGgn-0a}{\includegraphics[width=\scaleFactorA\textwidth]{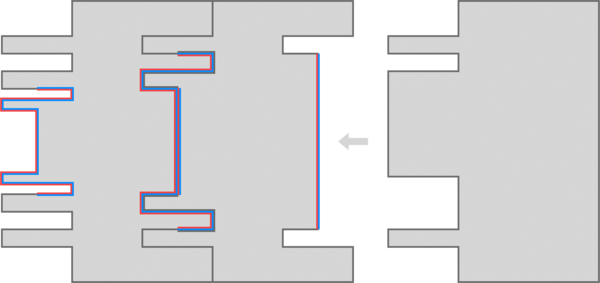}}}
\renewcommand{\figB}{\href{https://drive.google.com/open?id=1NfDwMGRBANZas6ZfBRV1NKbr44Gh8w7Y}{\includegraphics[width=\scaleFactorB\textwidth]{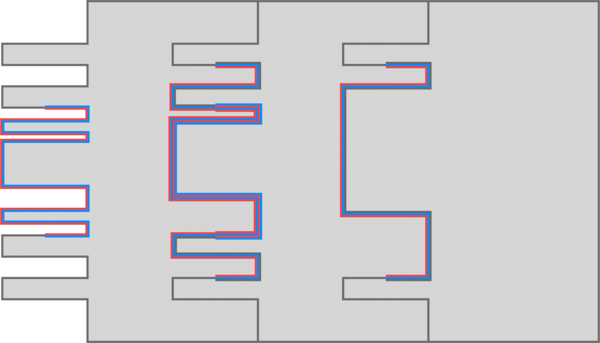}}}
\ifsmallFigures
\renewcommand{\scaleFactor}{0.25}
\vspace{3mm}
\centering
\tabulinesep=\tabcolsep
\begin{tabu} to \textwidth {|X[1,c,m]|X[1,c,m]|}
\hline
\figA & \figB \\
\hline
\multicolumn{2}{|c|}{\localcaption} \\
\hline
\end{tabu}
\else
\renewcommand{\scaleFactorA}{0.9}
\renewcommand{\scaleFactorB}{0.75}
\vspace{3mm}
\centering
\tabulinesep=\tabcolsep
\begin{tabu} to \textwidth {|X[1,c,m]|}
\hline
\figA \\
\figB \\
\hline
\localcaption \\
\hline
\end{tabu}
\fi
}

\caption{\sourcecodeQuoted{PolyCons (PolyCons Red)}}\label{FIG_--_M-PolyCons_PolyCons_Red}
\end{visupolfigure}



\notfinishedyet{\Cref{TODO} introduces \href{https://drive.google.com/open?id=1zRkvhtMlYzZ3bGtdNqwQNv7aI-CbyJK4}{\Mcons{Cons}}. To save space, we simply define a type-synonym to keep the number of \jointForm s in this article down:
\begin{sourcecodeEnv}
::List == SimpleType
\end{sourcecodeEnv}

}

\done{{| y2018_m08_d17_h16_m17_s09 |} move the following to the right position.}
\omitted{
\begin{visupolfigure}
\omitForInTestLatexBuilt{
\par
\vspace{1cm}
\newcolumntype{C}{>{\centering\arraybackslash} c }  
\centering
\begin{tabular}{C}
\href{https://drive.google.com/open?id=15jEZzspzZbWO3KFpuD7rfPe5VPlEI86y}{\includegraphics[width=0.8\textwidth]{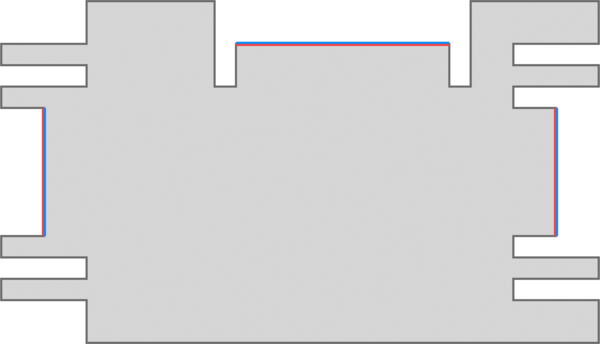}} \\
\Mcons{Cons} \\
\sourcecodeQuoted{List a <- a (List a)} \\
\end{tabular}
}
\caption{Overview of \mconstructor s.}\label{FIG_--_OverviewMconstructorsTwo}
\end{visupolfigure}
}

(\DemoExistingLaws.) \Cref{FIG_--_M-Cons_L_Cons_Red_E_R_L_Cons_L_Cons_E_E_R_R} shows a complex example of type-propagation. It shows that a ``list of lists'' is type-safe regarding its construction possibilities. Assume that the last \mconstructor\ that was joined, is the \mconstructor\ labelled with `1', an \href{https://drive.google.com/open?id=1oRpyLCpehG6z5gJQIXjnJoQL7LV-Ygii}{\Mcons{Red}}. It is joined with \mconstructor\ 2, an \href{https://drive.google.com/open?id=1zRkvhtMlYzZ3bGtdNqwQNv7aI-CbyJK4}{\Mcons{Cons}}. The male joint of 2 mimics the female joint that receives the \href{https://drive.google.com/open?id=1oRpyLCpehG6z5gJQIXjnJoQL7LV-Ygii}{\Mcons{Red}}. By simply obeying the laws of scaling en mimicking, it takes the form \href{https://drive.google.com/open?id=1sGTue01BHg0tuZ7SrGD24AIK7gDtYU6o}{\Mtype{List Colour}}. The female joint of 3 attached to the latter now takes the same shape. The other female joint of 3, by means of scaling and mimicking, now takes form \href{https://drive.google.com/open?id=1e65LShLcKxh4lT-7GvcO0lulG64ch7xS}{\Mtype{List (List Colour)}}. This propagates further to the female joints of 5 through 4. Note that in this process, by simply obeying the laws, the form scales up again, to take the form \Mtype{Colour} in the left female joint of 5.
\mustdo{{| y2018_m09_d05_h22_m13_s41 |} href to \Mtype{List Bool} as soon as this is created by jesse.}
The reader is invited to verify the following statements by visualisation:
\begin{enumerate*}[label=(\arabic*)]
   \item An \href{https://drive.google.com/open?id=1vYNs7Nsi-cdh3S9KopCQr4U_zgI4t0PQ}{\Mcons{True}} (see \cref{FIG_--_OverviewMconstructorsOne}) will not fit into the left female joint of 5. The same holds for a male joint with form \href{https://drive.google.com/open?id=1l67XMOveFw0ncfb1jwA-7KFbS82OVAH-}{\Mtype{List a}}, \Mtype{List Bool} and \href{https://drive.google.com/open?id=19id7w2rVu1Tr9p6eoz6X4feaRC2SzHmJ}{\Mtype{List (List Bool)}}, or any other form that is not \Mtype{Bool}.
   \item One can remove \href{https://drive.google.com/open?id=1oRpyLCpehG6z5gJQIXjnJoQL7LV-Ygii}{\Mcons{Red}}, and replace it with an \href{https://drive.google.com/open?id=1vYNs7Nsi-cdh3S9KopCQr4U_zgI4t0PQ}{\Mcons{True}}.
   \item If the latter point has been carried out, an \href{https://drive.google.com/open?id=1vYNs7Nsi-cdh3S9KopCQr4U_zgI4t0PQ}{\Mcons{True}} will fit into the left female joint of 5. An \href{https://drive.google.com/open?id=1oRpyLCpehG6z5gJQIXjnJoQL7LV-Ygii}{\Mcons{Red}} will not fit anymore.
   \item\label{LIST_--_Red-replaced-by-Cons} The \href{https://drive.google.com/open?id=1oRpyLCpehG6z5gJQIXjnJoQL7LV-Ygii}{\Mcons{Red}} can be replaced with an \href{https://drive.google.com/open?id=1zRkvhtMlYzZ3bGtdNqwQNv7aI-CbyJK4}{\Mcons{Cons}}.
   \item If point \labelcref{LIST_--_Red-replaced-by-Cons} has been carried out, the male joints of 3 and 4 will take the form \Mtype{List (List (List a)))}.
   \item If point \labelcref{LIST_--_Red-replaced-by-Cons} has been carried out, a female joint with form \Mtype{List (List (List (List Colour)))} will fit into the male joint of 3.
   \item If the latter point has been carried out, the female joint of 2 will take the form \href{https://drive.google.com/open?id=1e65LShLcKxh4lT-7GvcO0lulG64ch7xS}{\Mtype{List (List Colour)}}.
\end{enumerate*}

\begin{visupolfigure}
\omitForInTestLatexBuilt{
\par
\vspace{1cm}
\center
\href{https://drive.google.com/open?id=15jEZzspzZbWO3KFpuD7rfPe5VPlEI86y}{\includegraphics[width=1\textwidth]{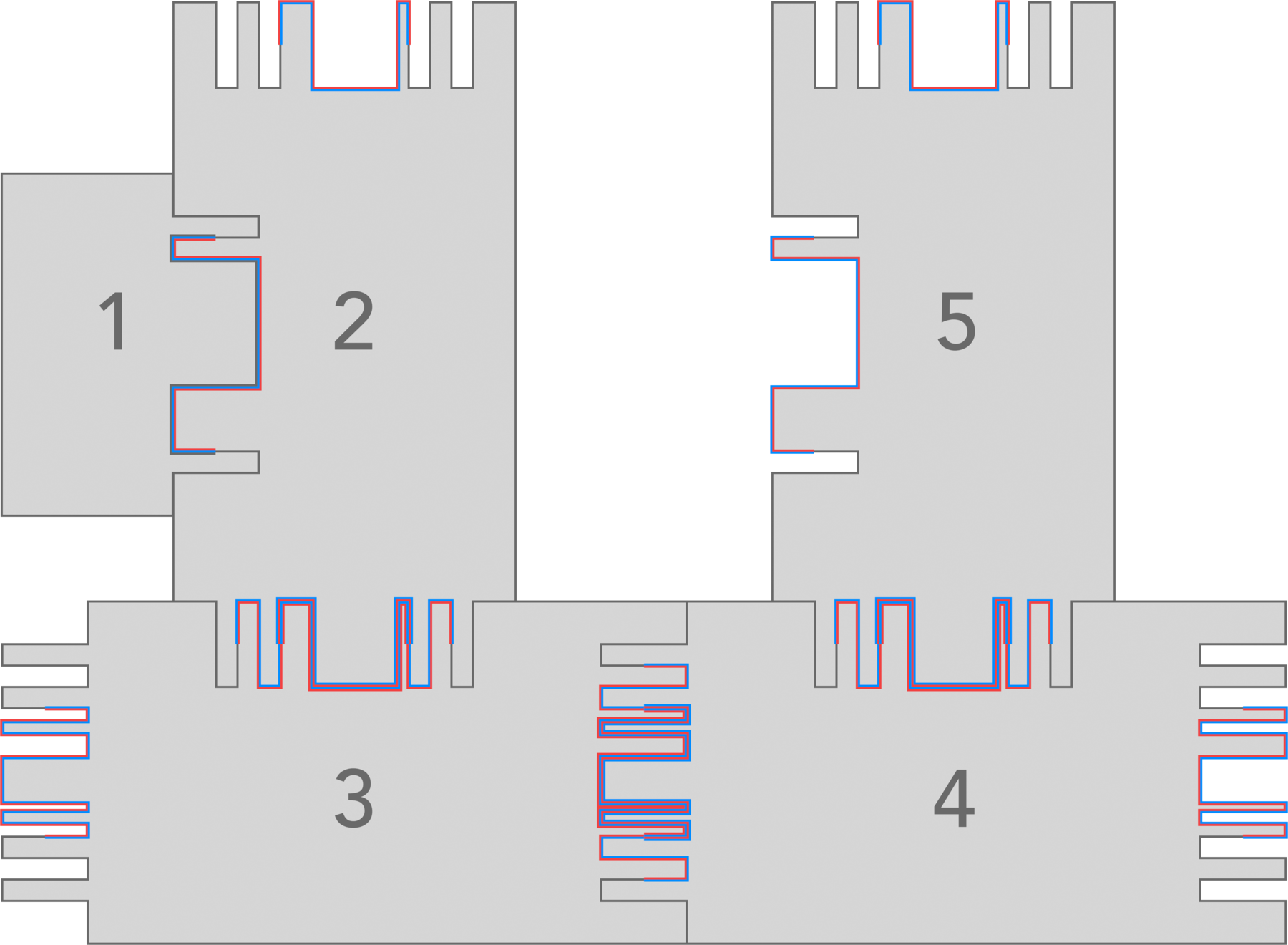}}
}
\caption{\sourcecodeQuoted{Cons (Cons Red \_) (Cons (Cons \_ \_) \_)}\omitted{ (List[List[Red],[\_]])}}\label{FIG_--_M-Cons_L_Cons_Red_E_R_L_Cons_L_Cons_E_E_R_R}
\end{visupolfigure}


The final example, \cref{FIG_--_True-into-a} shows a cross-section of a full 3D example. It uses the original \typeConsForm s (\cref{SEC__MadawipolA-by-Example}).

\begin{visupolfigure}
\omitForInTestLatexBuilt{
\begin{center}
\href{https://drive.google.com/open?id=1V_kUIqJXUeBe8sG8Pg0kExA44UfsG85i}{\includegraphics[width=0.45\textwidth]{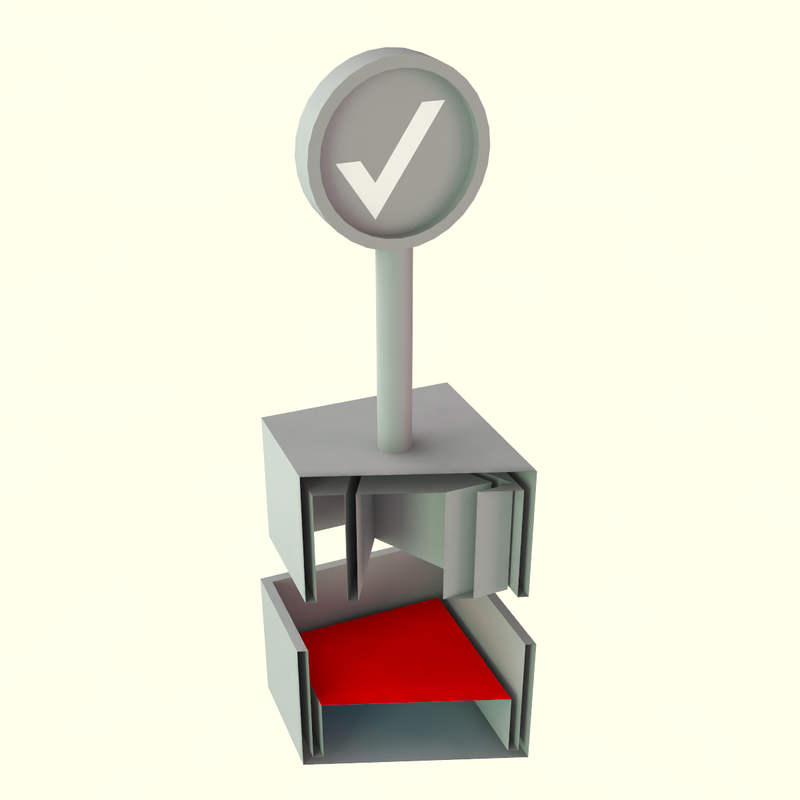}}
\href{https://drive.google.com/open?id=15SMDfVxhWVrH-S3H7hOH_k5ow3IOpOsT}{\includegraphics[width=0.45\textwidth]{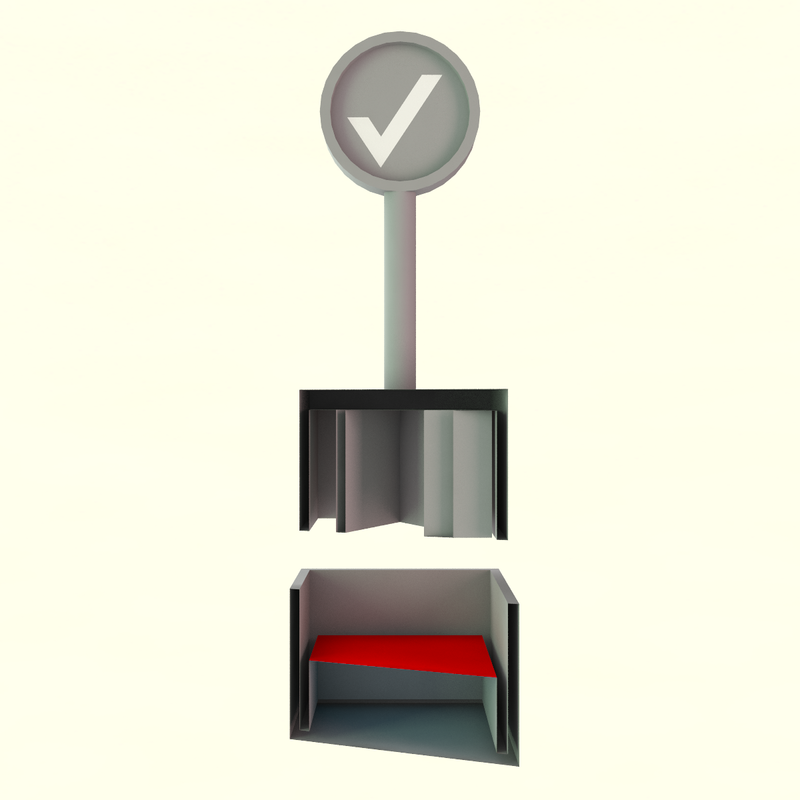}}
\par
\vspace{1cm}
\href{https://drive.google.com/open?id=1dBAdgxXCGqLkIXH8e1wfjOH8SlPh7VPf}{\includegraphics[width=0.45\textwidth]{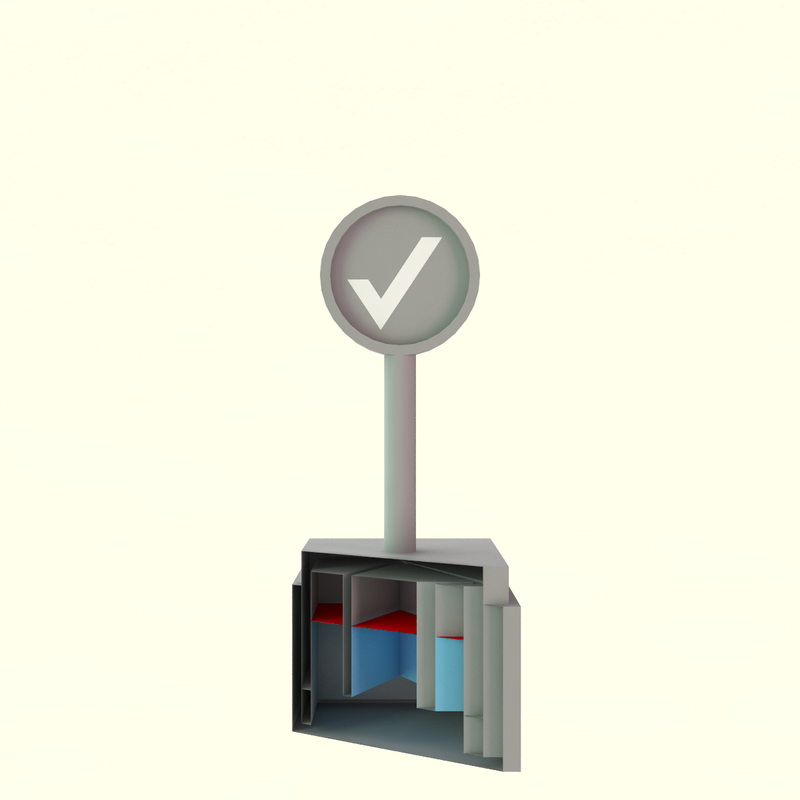}}
\href{https://drive.google.com/open?id=1sUkaG9pWVUb1KqbgagbUu-2rjqXAswrB}{\includegraphics[width=0.45\textwidth]{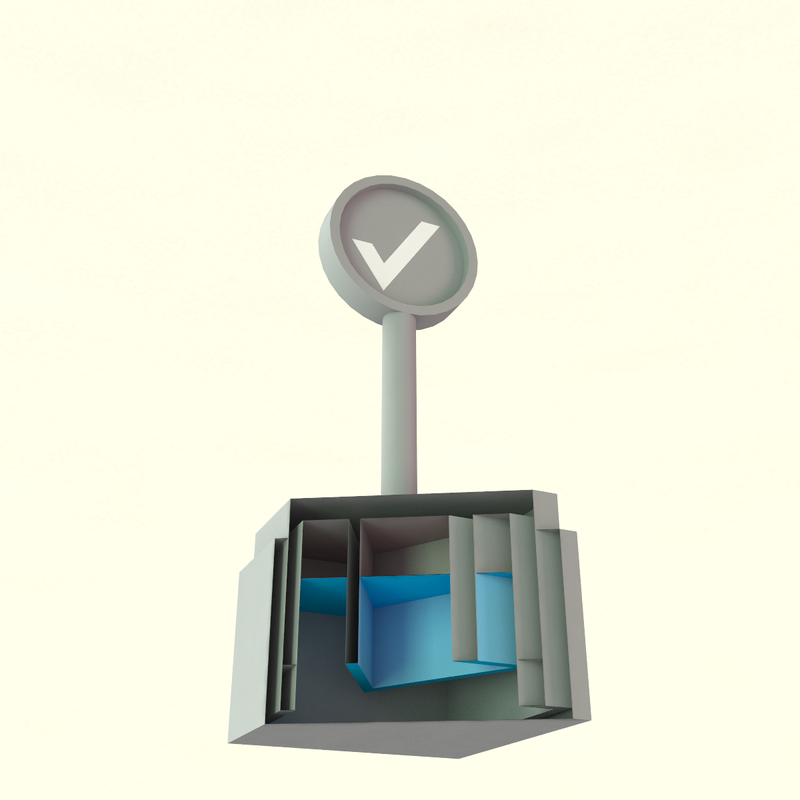}}
\end{center}
}
\caption{\Mcons{True} into \Mtype{a}.\omitted{ The \mconstructor s are presented as cross-sections. The upper row shows two perspectives before insertion, and the lower row shows two perspectives after half-way insertion.}}\label{FIG_--_True-into-a}
\end{visupolfigure}


\mustdo{{| y2018_m05_d15_h18_m41_s22 |} somewhere explain why the polymorphic surfaces are halfway, using examples in which they are placed somewhere else!}

\notfinishedyet{}
{  
\mustdo{{| y2018_m05_d11_h16_m31_s01 |} use the following in some section.}
\begin{Law}[\MimickingPolSurf]
In general, all \polymorphicSurface s that are attached to the same object, will mimick each other's behaviour. Only one surface can be manipulated at a given time. A surface that is either in the state in which it is taking the shape of an object that is pushed into it, or that, is in state of \differentation-PolSurf. \MimickingPolSurf\ \PolymorphicSurface s that are in a state of \differentiation\ will not take the shape of objects that are moved into them.
\end{Law}
}

\mustdo{{| y2018_m05_d12_h14_m52_s31 |} create a minimal example that shows type information propagation.}
\notfinishedyet{}
{
\subsection{Type propagation}
}

\section{Definition and correctness of \madawipolB}\label{SEC__Definition-MadawipolB}

This section defines essential aspects of \madawipolB. It does so in a semi-formal way, both to save space, and to be more legible for a wider audience. \cite{groenouwe2017} contains a more formal definition of \madawipolA. There is a great degree of overlap between \madawipolA\ and \madawipolB. A reader that appreciates additional precision, may therefore be considerably served by the latter article.

The syntax and semantics of \madawipolB\ is defined by means of a translation between \madawipolB\ and the textual language (\adtpolyOne). This way, the already existing semantics of the textual language is transferred to \madawipolB. Note, however, that a mere definition is not sufficient. \madawipolB\ should also exhibit the same relevant behaviours as its textual counterparts: type-safety. This article suffices to support this with the intuitive examples that have been provided so far. \cite{groenouwe2017} provides a mathematical proof of, among other things, type-safety of \madawipolA. This proof could serve as an inspiration for a proof for \madawipolB. However, this is beyond this article's scope. Nevertheless, \cref{SEC__-Proof} briefly reflects on a proof strategy.

\omitted{This is sufficient for defining the syntax: the latter is simply equal to the range of $\trans$. However, it is not sufficient for defining \madawipolA's semantics. A translation does not yet \emph{prove} that the target language exhibits the same relevant behaviours as the original language. This article, however, will not include a formal proof.}

\mustdo{{| y2018_m08_d24_h18_m11_s24 |} place the section below in the right place.}

\subsection{The \transConfig}

The translation from \adtpolyOne\ and \madawipolB\ is defined recursively on the structure of \adtpolyOne. For this purpose, predefined `atomic' translation information from \adtpolyOne\ has to be supplied to this translation. This information is provided in the \newterm{\transConfig}.

The \transConfig\ consists of the following elements:
\begin{enumerate}
   \item A set of \ADTD s $\FsetADTD$.
   \item The \newterm{\typeConsMap}: A mapping from each \typeCons\ occurring in $\FsetADTD$ to a \typeConsForm.
   \item The \newterm{\constrBlockMap}:  A mapping that maps a given constructor occurs in $\FsetADTD$ with a solid 3D object that is the \mconstructor\ in its rough form: it does not yet contain the joints.
   \item The \newterm{\constrArgsMap}: A mapping from each constructor $\FsetADTD$ and an argument position, to the location and orientation of the joint that corresponds with the textual argument position. (For example, it would map the first argument-position of \sourcecodeQuoted{Cons} to the location and orientation of the joint on the top in \cref{FIG_--_OverviewMconstructorsOne}.)
   \item The \newterm{\constrResultTypeMap}: A mapping from each constructor in $\FsetADTD$ to the location and orientation of the male joint, hence the result type of the constructor.
\end{enumerate}
\mustdo{{| y2018_m08_d21_h00_m58_s42 |} write, transfer from visupol2016 article.}
The \transConfig\ should comply with a number of conditions to be valid. Some of these are covered in the rest of this section.

The rest of this section explains some core concepts of the \transConfig\ more precisely. One of these is the way types expressed in \adtpolyOne\ are mapped to \typeForm s.\coulddo{ This is the topic of \cref{SEC__Definition-MadawipolB}.}\mustdo{{| y2018_m08_d26_h11_m50_s31 |} complete this overview.}

\subsection{\TypeConsForm s}\label{SEC__Definition-MadawipolB__typeConsForms}

A more precise definition of \typeConsForm s follows now. We revisit \cref{FIG_--_Several-typeConsForms}: it provides a list of examples, when reinterpreting them as 2D projections of the joints perpendicularly to the bottom of the joint. The outer squares in this figure, the \alignmentSquare\ is not part of the \typeConsForm.\shoulddo{{| y2018_m08_d21_h00_m34_s11 |} provide example of what could go wrong if you do not have such a joint.}\done{{| y2018_m08_d21_h00_m35_s18 |} is this the right place to introduce the alignment square, should probably have been done earlier.}
\done{{| y2018_m08_d23_h16_m37_s54 |} perhaps rename polymorphicSubspace simply to PolymorphicSurface, to cut down on terminology. [{| y2018_m08_d23_h19_m16_s38 |} no can't be done: the polymorphicSubspace consists of a PolymorphicSurface and a linear transformation, so you need both terms.]}

Let $T$ be a \typeCons. The \typeConsForm\ $T_f$ of $T$ is a pair consisting of a \rigidPart\ $T_R$ and, if applicable, a \polymorphicSubspace\ $T_S$: $T_f = (T_R, T_S)$. $T_R$ is a surface in 2D space. In \cref{FIG_--_Several-typeConsForms} all gray\mustdo{{| y2018_m08_d22_h20_m38_s37 |} adapt once Jesse has provided the final figures.} forms with exclusion of the \alignmentSquare s are examples. For referencing purposes, assume that the origin of this plane is in the center of the \alignmentSquare.

If $T$ has a \typeParameter, then $T_f$ has a \polymorphicSubspace. This is a pair that consists of a \polymorphicSurface\ $T_P$ and a \tcArgumentTransformation\ $T_A$. Hence, $T_S = \textsf{Optional}[(T_P, T_A)]$\footnote{$\textsf{Optional}$ is as \sourcecodeQuoted{Maybe} in Haskell}. $T_P$ is a surface in 2D space, too. In \cref{FIG_--_Several-typeConsForms} these are indicated with a red colour. The figure contains an example that contains a \polymorphicSubspace: \href{https://drive.google.com/open?id=1l67XMOveFw0ncfb1jwA-7KFbS82OVAH-}{\sourcecodeQuoted{List a}}. \mustdo{for the case the permittedZone after scaling, translation and rotation is within the polymorphic subspace (and not coinciding with it): (The situation can be somewhat more involved with other forms, this will be explained in ...) }\mustdo{{| y2018_m08_d21_h17_m41_s04 |} in a black and white paper, you cannot use colours, so revise in one way or the other.}

The \tcArgumentTransformation\ $T_A$ is a linear transformation. This transformation is, loosely speaking, used to scale down the argument of a maramafied \typeCons, so that it fits within its \polymorphicSubspace. In \cref{FIG_--_Several-typeConsForms} this \tcArgumentTransformation\ is simply indicated with a green square. In the case of \href{https://drive.google.com/open?id=1l67XMOveFw0ncfb1jwA-7KFbS82OVAH-}{\sourcecodeQuoted{List a}}, the green square coincides with the edge of the \polymorphicSubspace. A precise explanation of \tcArgumentTransformation s follows in \mustdo{TODO}.

If $T$ does not have a \typeParameter, then it does not have a \polymorphicSubspace\ (hence, $T_S$ is $\textsf{None}$).

\subsection{\TypeForm s}\label{SEC__Definition-MadawipolB__typeForms}

\done{{| y2018_m08_d21_h12_m48_s28 |} is the alignment square included in the typeForm and in the typeConsForm? (answer is: not in the typeConsForm, but it is in the typeForm.}
Each \ADT\ has a corresponding \typeForm\ in \madawipolA. In fact, we have already seen some of these in \cref{FIG_--_Several-typeConsForms}: these are all examples of \typeForm s of types that contain exactly one \typeCons. A \typeForm\ also consists of a pair, that contains a \rigidPart\ and, possibly, a \polymorphicSubspace.

Examples of \typeForm s of more complex types, such as \href{https://drive.google.com/open?id=19id7w2rVu1Tr9p6eoz6X4feaRC2SzHmJ}{\Mtype{List (List Bool)}}, are provided in \cref{FIG_--_Several-typeForms}.

\begin{visupolfigure}
\omitForInTestLatexBuilt{
\renewcommand{\scaleFactor}{0.30}
\par
\vspace{1cm}
\newcolumntype{C}{>{\centering\arraybackslash} c }  
\centering
\begin{tabular}{CCC}
\href{https://drive.google.com/open?id=1sGTue01BHg0tuZ7SrGD24AIK7gDtYU6o}{\includegraphics[width=\scaleFactor\textwidth]{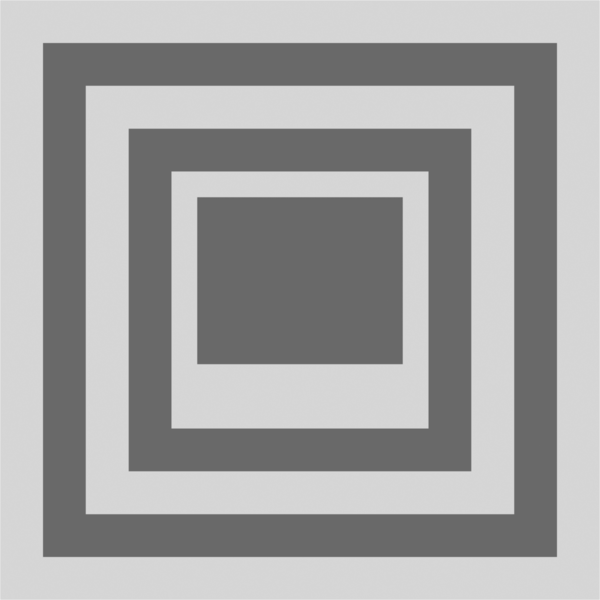}} & %
\href{https://drive.google.com/open?id=1e65LShLcKxh4lT-7GvcO0lulG64ch7xS}{\includegraphics[width=\scaleFactor\textwidth]{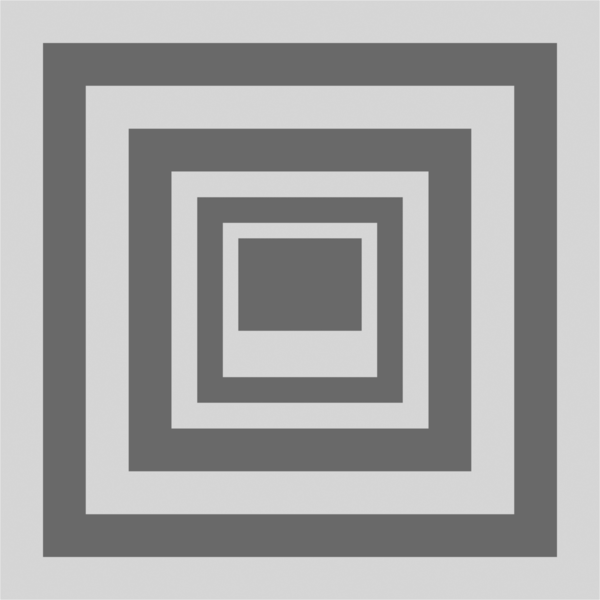}} & %
\href{https://drive.google.com/open?id=19id7w2rVu1Tr9p6eoz6X4feaRC2SzHmJ}{\includegraphics[width=\scaleFactor\textwidth]{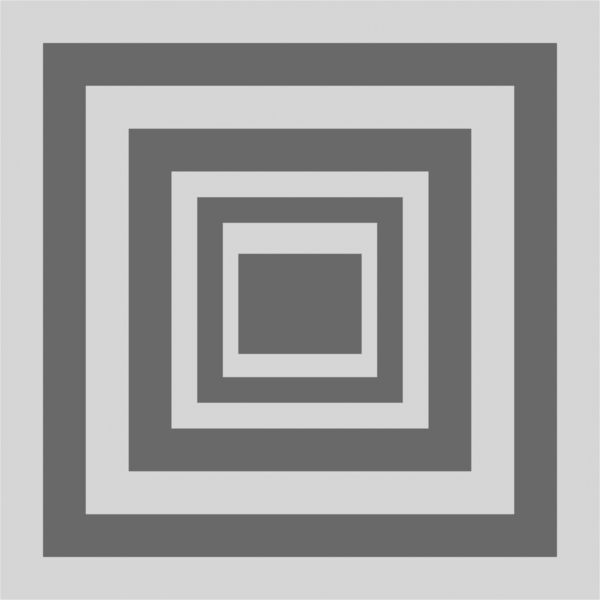}} \\ %
\Mtype{List Colour} & \Mtype{List (List Colour)} & \Mtype{List (List Bool)} \\
\href{https://drive.google.com/open?id=1hOt8s2Igb5Q7_yNQgrEWv1AxJuQR4GjU}{\includegraphics[width=\scaleFactor\textwidth]{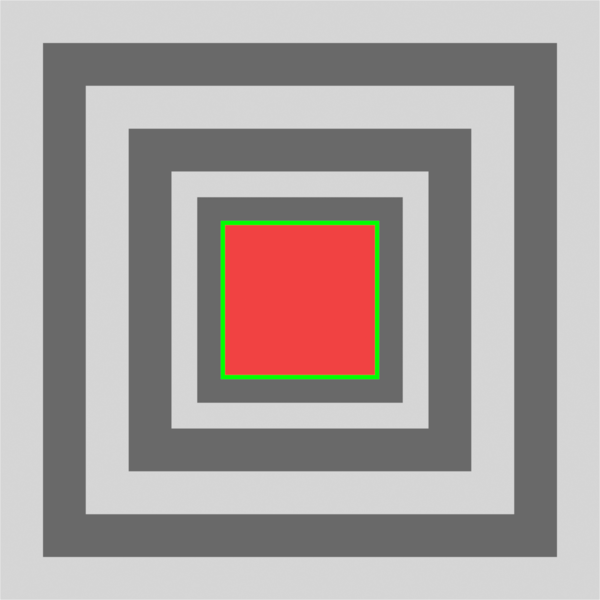}} &
 & 
\\
\Mtype{List (List a)} & & \\ 
\end{tabular}
}
\caption{Several \typeForm s.}\label{FIG_--_Several-typeForms}\shoulddo{{| y2018_m08_d21_h00_m30_s46 |} also add the side cross-section of each joint, and perhaps the joint in 3D.}
\end{visupolfigure}

Such complex \typeForm s can be obtained by means of an inductively defined procedure applied to the \typeCons s and \typeParameter s that occur in the type.\shoulddo{Note that in the actual editor, such \typeForm s are often not created directly, but indirectly by type-propagation, as will be shown later in this article.} For the procedure, first two definitions are needed that define how to scale down an argument of a maramafied \typeCons.
\begin{definition}[\tcArgumentTransformation\ application]
Let $S$ be a subset of the 2D plane, and $L$ a 2D linear transformation. Then $L \cdot S$ is the transformation of $S$ with $L$, i.e.~the set of all points in $S$, interpreted as a set of 2D vectors, transformed by means of $L$. Moreover, let $L'$ be another 2D linear transformation. Then $L \cdot L'$ is the transformation that is obtained by composing $L$ and $L'$, hence by first applying $L'$ and then applying $L$. This is all `just' standard linear algebra \cite{mirsky2012introductionLinearAlgebra}.\omitted{, in particular if one represents the \tcArgumentTransformation\ as a matrix, matrix multiplication.}
\end{definition}

\begin{definition}[\tcArgumentTransformation\ application to  a \typeCons]
Let $L$ be a linear transformation and $T_f$ a \typeForm. Then $L$ applied to $T_f$ is defined as applying $L$ to all elements of $T_f$. More precisely, let $T_f = (T_R, (T_P,T_A))$, where $T_R$ is the \rigidPart\ of $T_f$, $T_P$ its \polymorphicSurface, and $T_A$ its \tcArgumentTransformation. Then $L \cdot T_f = (L \cdot T_P, (L \cdot T_R, L \cdot T_A))$.
\end{definition}
\newcommand{\typeFormProcedure}{$\textsf{typeFormProcedure}$}

These definitions allows the definition of \typeFormProcedure:

\begin{definition}[\typeFormProcedure]
The procedure \typeFormProcedure\ determines the \typeForm\ for a given \ADT. In words, it is defined as follows. The definition uses \href{https://drive.google.com/open?id=1hOt8s2Igb5Q7_yNQgrEWv1AxJuQR4GjU}{\sourcecodeQuoted{List (List a)}} as a running example.
\begin{itemize}
   \item Let $T$ be the outermost \typeCons, and $T_f = (T_R, (T_P,T_A))$ its \typeConsForm. (Example: recursive pass 1: $T$ is \sourcecodeQuoted{List}, recursive pass 2: $T$ = \sourcecodeQuoted{List}.)
   \item If $T$ does not have an argument, or if it has an argument that is a \typeParameter, then return $T_f$. (Example: the condition only holds for pass 2: $T$ has argument \sourcecodeQuoted{a}.)
   \item Otherwise, let $T'$ be the argument of $T$. Note that $T'$ contains at least one \typeCons. Now apply recursion: apply \typeFormProcedure\ to $T'$, and let $T'_f$ be the resulting \typeForm. (Example: the condition only holds for pass 1: $T'$ = \sourcecodeQuoted{List a} and $T'_f$ is \Mtype{List} (See \cref{FIG_--_Several-typeConsForms})).)
   \item Now apply the \tcArgumentTransformation\ of $T_f$ to $T'_f$. Then, add the \rigidPart\ of $T_f$ to the \rigidPart\ of the latter application. Return the result.
   An alternative precise formulation: return $(T_R \cup (T_A \cdot T'_R), (T_A \cdot T'_P, T_A \cdot T'_A))$. (Example: only applies to pass 1: the result is the form in \cref{FIG_--_Several-typeForms} with \Mtype{List (List a))} underneath.)
\end{itemize}
\end{definition}
The reader is invited to verify that the \typeForm s in \cref{FIG_--_Several-typeForms} are indeed constructed in accordance with \typeFormProcedure, by applying the procedure to the types underneath each form.\omitted{Note, in is the result of two consecutive transformations on the \typeConsForm\ of \Mtype{Bool}: one as a consequence of the \tcArgumentTransformation\ associated with the \typeConsForm\ of the innermost \Mtype{List}, and the second as a consequence of the \tcArgumentTransformation\ associated with the \typeConsForm\ of that surrounds it, in this case again a \typeConsForm\ of \Mtype{List}.}
\mustdo{{| y2018_m08_d21_h22_m40_s39 |} add polymorphicSubspace!}

Note that the original type can be immediately recognised in the \typeForm\ by starting with the outermost \typeConsForm\ and then walking your way to the adjacent one, all the way down to the center form, while reading out loud the associated \typeCons s. The reader is invited to verify that, in the examples in \cref{FIG_--_Several-typeForms}, that indeed leads to reading out loud the types written underneath each form.

Also, it may be clear that due to its recursive nature, this procedure defines a transformation of any type into a \typeForm.

\subsection{\MaleJntForm}

In precise terms, the 3D \maleJntForm\ for a \typeForm\ is obtained as follows. The \rigidPart\ of the \maleJntForm\ is obtained as follows. First add an \alignmentSquare\ to the \typeForm's \rigidPart. Then add a third spatial axis to the \typeForm's space, turning it into a 3D space, and extrude the \rigidPart\ of the \typeForm\ vertically upwards over a distance of $\FverticalJointSize$ (the  \verticalJointSize).\mustdo{{| y2018_m08_d24_h14_m28_s49 |} introduce verticalJointSize} The front side of the joint is defined as the part before extruding began, so the part in the plane $z = 0$. In other words, if the joint is joined, it is this part that enters the opposing joint first. Hence, note that when facing the joint from the front, it will be a mirror image of the \typeForm. Complete the \rigidPart\ by adding a bottom to the joint by adding a square surface that covers the joint's complete bottom (so, at height $z = \FverticalJointSize$).

The \polymorphicSurface\ of the joint is obtained, by first painting the bottom side of \polymorphicSurface\ red, and the top side blue, and then moving (not extruding) the \polymorphicSurface\ of the form upwards over a distance of $\frac{1}{2} \FverticalJointSize$.

\subsection{\FemaleJntForm}

The 3D \femaleJntForm\ for a \typeForm\ $T_f$ is obtained as follows. The \rigidPart\ is obtained as follows. First take the union $U$ of an \alignmentSquare, $T_f$'s \rigidPart\ and $T_f$'s \polymorphicSurface. ($T_f$'s \polymorphicSurface\ is included, because this creates space for the placement of the joint's \polymorphicSurface\ in the next step.) Then take the complement of $U$ enclosed within the outer edge of the \alignmentSquare, leading to $\overline{U}$. Then extrude $\overline{U}$ vertically downwards over a distance of $\FverticalJointSize$. Then add a bottom to the joint by adding a square surface that covers the joint's complete bottom (so, at height $z = -\FverticalJointSize$). The front side of the joint is, again, defined as the part before the extruding began. Note that the \femaleJntForm\ aligns exactly with the \typeConsForm. It is not a mirror image.

The joint's \polymorphicSurface\ is obtained by first painting its top blue and its bottom red. Move (so, not extrude) $T_f$'s \polymorphicSurface\ downwards over a distance of $\frac{1}{2} \FverticalJointSize$. Then, extrude only the edge of this \polymorphicSurface\ further downwards over the rest of the distance of the joint, so another $\frac{1}{2} \FverticalJointSize$. The latter is needed to attach the \polymorphicSurface\ to the joint's bottom, otherwise it would be free-floating. The material created by extruding this edge is rigid: it is not part of the joint's \polymorphicSurface.

\shoulddo{{| y2018_m08_d21_h12_m32_s16 |} create and include pictures of this process.}

\subsection{Condition \typeConsForm s}\label{SEC__Definition-MadawipolB__condition-typeconsforms}

A condition that \typeConsForm s have to meet is that each pair of distinct \typeConsForm s, should be such that their corresponding \maleJntForm\ and \femaleJntForm\ do not fit into each other. The following expresses this condition in a precise and operational form.

Let $T_f$ be a \typeForm\ or \typeConsForm\ . The \newterm{\femaleBottomRegion} of $T_f$ is the region of $T_f$ that in the corresponding \femaleJntForm\ is at the height of the bottom of the joint, or can go down to that level. In precise terms, it is equal to the union of $T_f$'s  \rigidPart\ and its \polymorphicSurface, with exception of the edge of this \polymorphicSurface. For example, in \href{https://drive.google.com/open?id=1l67XMOveFw0ncfb1jwA-7KFbS82OVAH-}{\Mtype{List a}}'s form in \cref{FIG_--_Several-typeConsForms} it is equal to the inner thick square plus the inner square region, minus the edge of the inner square region.

Let $T_{f}$ and $T'_{f}$ be two \typeConsForm s. One can prove that the following holds: a male joint with form $T_{f}$ will not fit into a female joint of $T'_f$ if, and only if, the \rigidPart\ of $T_{f}$ contains points that are not in \femaleBottomRegion\ of $T'_{f}$.

The condition can now be formulated as that for any pair of \typeConsForm s present in the \transConfig, the \rigidPart\ of the first contains points that are not in the \femaleBottomRegion\ of the second, and vice versa: the \rigidPart\ of the second contains points that are not in the \femaleBottomRegion\ of the first.

\mustdo{{| y2018_m08_d26_h13_m42_s02 |} move to right spot, this section is about defining, not proving: It is then also possible to prove mathematically that only the \typeForm s of types that are identical, or one is a supertype of the other fit. In other cases they will not fit.}

\mustdo{{| y2018_m05_d15_h18_m07_s41 |} also explain somewhere in this section why the polymorphic surface is located half way the joint, and the reason that the law of these surfaces is that they move in two direction at the same time.}

\subsection{Translation}

\madawipolB's expressivity is limited to \ads s. Therefore, the translation $\trans$ from textual language (\adtpolyOne) to \madawipolB\ consists of a translation between \ads s. These \ads s may be unfinished (see \cref{EX_--_-Unfinished-ADS}). It can be defined recursively on the structure of \ads s. $\trans$ takes two arguments, an \ads\ $\Fads$ from \adtpolyOne\ and a \transConfig\ $\FtransConfig$, hence an application can be written as $\trans(\Fads, \FtransConfig)$. It is, sketch-wise, defined as follows. If $\Fads$ is atomic, then use the information expressed in $\FtransConfig$ to build the corresponding \mconstructor. Otherwise, translate the top-level arguments of $\Fads$, and join these with the corresponding joints of the translation of the outermost constructor of $\Fads$. If at any moment in this process, some joints cannot be joined, then the translation results in $\Funjoinable$. The technical description of the translation is beyond the scope of this article.

\section{Proof of semantical equivalence of \madawipolB\ and \adtpolyOne}\label{SEC__-Proof}

Just as \adtpolyOne, \madawipolB, should be type-safe. This may now be intuitively clear, based on the examples developed so far.

A proof is beyond the scope of this article, however, some reflections on it follow now. Type-safety is equivalent to stating that $\trans$ never maps a well-typed \ads\ to $\Funjoinable$, and conversely, that the inverse translation $\trans^{-1}$ of any \Mads\ from \madawipolB\ is well-typed in \adtpolyOne. Such a proof has been provided for \madawipolA\ \cite{groenouwe2017}. In \madawipolB\ there is an extra challenge because of type-propagation. A proof strategy is to interpret each \MadsE\ as a typing statement. Joining of \mconstructor s, then, can be regarded as a form of typing derivation, and, hence, as part of a type-system \cite[Sec.~8]{pierce2002types}. After all, by reading the types that occur in the joints after \mconstructor s are joined, one has derived the type. This means that the proof of type-safety can be realised by showing that the latter type system is isomorphic to the type system of \adtpolyOne. In other words, for each typing derivation rule of the type system of \adtpolyOne\ one has to define a counterpart based on joining \mconstructor s from \madawipolB, and then one has to prove that these behave congruently.

\clog{
[{| y2018_m08_d25_h19_m09_s48 |} Comparing the proof and approach in madawipolA and madawipolB. In the first, I first define the set of all possible mconstructors, by translating all constructors with a closed type to madawipolA. Then I deine translation from text to madawipolA, by first type-annotating a given ADS. If this ADS is a literal (a `stand-alone' value), then all polymormpy is resolved - I.e.~all constructors are instantiated to a closed type. Then translate each individual constructor to its corresponding mconstructor in the set I just mentioned.

In madawipolB another approach is needed, because the set of mconstructors do not have to contain closed-type instantiations of mconstructors. For example, it only has to contain Cons:List a <- a (List a), and for example not Cons:List Bool <- Bool (List Bool). The translation is much easier, I can skip the type-annotation of a given ADS from the textual language, and just translate each constructor into its corresponding mconstructor using the most general type it can take based on its ADT definition.
}

\subsection{Unifiable = joinable}

An important part of the type-safety of \madawipolB\ leans on the fact that any pair of types is unifiable if, and only if the corresponding male and female \jointForm s fit into each other, assuming that both forms were undifferentiated before the fitting started.\shoulddo{{| y2018_m09_d02_h19_m53_s15 |} brief definition of unification.}\mustdo{cite definition of unifiable.}
In other words: ``unifiable = joinable''. This is a central lemma in a proof of type-safety.  To build some intuition that this lemma is true, the reader could try to verify this for the following statements, by using \cref{FIG_--_Several-typeForms}. Hence, investigate for each of the following statements whether the male joint with the first \typeForm, joins with the female joint with the second \typeForm, and whether the male joint with the second \typeForm, joins with the female joint with the first \typeForm\ if, and only if the types are unifiable. Whether they are is indicated with $\FunifiableWith$ (types are unifiable) or $\FnotUnifiableWith$ (they are not).
\href{https://drive.google.com/open?id=1l67XMOveFw0ncfb1jwA-7KFbS82OVAH-}{\Mtype{List a}} $\FunifiableWith$ \Mtype{a}%
; \href{https://drive.google.com/open?id=1hOt8s2Igb5Q7_yNQgrEWv1AxJuQR4GjU}{\Mtype{List (List a)}} $\FnotUnifiableWith$ \Mtype{List Bool}%
; \href{https://drive.google.com/open?id=19id7w2rVu1Tr9p6eoz6X4feaRC2SzHmJ}{\Mtype{List (List Bool)}} $\FnotUnifiableWith$ \href{https://drive.google.com/open?id=1e65LShLcKxh4lT-7GvcO0lulG64ch7xS}{\Mtype{List (List Colour)}}%
; \href{https://drive.google.com/open?id=1l67XMOveFw0ncfb1jwA-7KFbS82OVAH-}{\Mtype{List a}} $\FunifiableWith$ \href{https://drive.google.com/open?id=1hOt8s2Igb5Q7_yNQgrEWv1AxJuQR4GjU}{\Mtype{List (List a)}}%
.
\coulddo{{| y2018_m09_d02_h20_m14_s52 |} the examples would be better if there would be a second typeCons that takes an argument, we now only have List... For example, introduce `Pair'.}

\cite[Lemma~2, p.~11]{groenouwe2017} is a proof of this lemma for \madawipolA. A proof for \madawipolB\ is beyond this article's scope. However, it is very similar. A proof-sketch follows now. If $T$ and $T'$ are unifiable, then either $T = T'$, and then the joints trivially fit, or $T$ and $T'$ consist of a sequence of \typeCons-applications that have an identical beginning sequence, starting from the outermost \typeCons. After this beginning sequence, either $T$ or $T'$ ends in a \typeParameter. Assume it is $T$ that does so. Example: $T$=\sourcecodeQuoted{List (Pair (List a))} and $T'$=\sourcecodeQuoted{List (Pair (List (List Bool)))}. Here, the identical beginning sequence is \scq{List (Pair (List ...}, after which $T$ ends in the \typeParameter\ \scq{a}, while $T'$ contains additional \typeCons s. The part of the \jointForm s that are a result of the translation of this beginning sequence will always fit. One can easily see this by investigating how \typeFormProcedure\ works: each \typeConsForm\ is only influenced by the translation of the preceding \typeCons s, never the later. The rest of the joints also fit. The \typeParameter\ of $T$ is translated into a \polymorphicSurface. Due to the linear transformation it went through, it will completely cover the translation of the remaining sequence of \typeCons s of $T'$ (in the example: \scq{List Bool}). Hence, these parts of the joints will also fit. It is easy to see that this works both for a male version of $T$ into the female version of $T'$, and the male version of $T'$ into the female version of $T$.

If $T$ and $T'$ are not unifiable, then there must be a minimal position $n$ where the \typeCons\ of $T$ is different from that of $T'$. For example, with $T$=\Mtype{List (Pair (Pair a))} and $T'$=\sourcecodeQuoted{List (Pair (List (List Bool)))} this is the case at the $3_{rd}$ position, so $n = 3$. The corresponding \typeConsForm s at position $n$ of both joints is linearly transformed in exactly the same way by \typeFormProcedure, because the sequence of \typeCons s that precedes them is identical. Joints with different \typeConsForm s do not fit into each other. This is a requirement of the \transConfig\ (see \cref{SEC__Definition-MadawipolB__condition-typeconsforms}). Obviously, they also do not fit after going through the same linear transformation. Consequently, the joints do not fit.

A formal proof of the previous sketch can easily be obtained by induction.

\section{Related work}

The scientific literature on designs related to capturing polymorphic data structures, in particular \adsE\ with \typeParameter s is to the best of our knowledge confined to \cite{Lerner2015polymorphicBlocks}. They created a design based on a similar idea of which we became aware retrospectively: \polyBlocks\ (\polyBlocksA). Although the ideas are similar, the designs of \madawipolB\ has the following advantages over \polyBlocksA:
\begin{enumerate}
   \item \madawipolB\ is more expressive (except for its current limitation to one \typeParameter, however, see \cref{SEC_--_Conclusion-and-future-work}). It essentially covers polymorphic \ADT s fully: unlimited type complexity (i.e.~types with any number of \typeCons s) and recursive type definitions in a well-typed way.  It does so by applying an effective scheme of linear transformations. \polyBlocksA\ also allows for some degree of type complexity, however, not fully. What is more, from the perspective of \ADT s, it is not type-safe. An example is the creation of a List of Lists: compare the visual \scq{Cons} in \cite[Fig.~3, $3^{rd}$ block]{Lerner2015polymorphicBlocks} and \cref{FIG_--_M-Cons_L_Cons_Red_E_R_L_Cons_L_Cons_E_E_R_R}: the \scq{Cons} of the first will not be able to transfer its type information, leading to ill-typed constructions. This can be clearly seen in \cite[Fig.~4, $3^{rd}$ block]{Lerner2015polymorphicBlocks}. The second argument of the \scq{Cons} is fully differentiated to a \scq{Bool} form. However, it is also compatible with a \scq{List} of \scq{Boolean}s. One could argue that this is a general form of type-unsafety, not only from the perspective of \ADT s.
   \item \madawipolB\ uses a 3D design, in which each visualised \typeCons\ encloses the visualised arguments. This may be more intuitive than the juxtaposition of visual \typeCons s in \polyBlocksA. Moreover, this 3D design supports the linear transformations of visual \typeCons-arguments needed for creating complex types more intuitively.
   \item \madawipolB\ introduces laws that explain why \polymorphicSurface s (called `polymorphic ports' in \polyBlocksA) behave as they do within constructors, while \polyBlocksA\ does not. (See \cref{SEC_--_Polymorphic-surfaces}). In particular, that female and male joints create `inverse' joint forms, is a consequence of simpler laws in \madawipolB. This may support intuitive understanding. 
\end{enumerate}

An advantage of \polyBlocksA\ over \madawipolB\ is that it supports multiple \typeParameter s, \madawipolB\ currently only one. \madawipolB, however, can easily and naturally be extended for this purpose.

\section{Conclusion and future work}\label{SEC_--_Conclusion-and-future-work}

This article presented the design of truly visual polymorphic and type-safe \adsE s as they occur in advanced languages such as Haskell and Clean. The design supports recursively defined types, and unlimited complex types consisting of more than one \typeCons. The type-safety is enforced solely mechanically, without requiring any text.

Future work on the design could include the following.
\begin{enumerate}
   \item Extend to multiple \typeParameter s. This is straightforward: allow more than one \polymorphicSurface to be present in a \typeForm, and give them distinct colours.
   \item \mconstructor\ s may physically get `into each other's way' depending on the shape of the \mconstructor s and the structure that one is trying to build (e.g.~try to build a \scq{List (List (List (List Bool)))}). Solve this problem.
   \item Empirically validate the understandability of the design.
\end{enumerate}

\section*{Acknowledgments}

We express our gratitude towards Douwe Schulte for his assistance in editing the paper.

This work is part of the STW research programme with project number 13855, which is (partly) financed by the Netherlands Organisation for Scientific Research (NWO).
\coulddo{{| y2018_m09_d06_h22_m18_s09 |}}

\bibliographystyle{./imported_latex_style_files/eptcs/eptcs}
\bibliography{bib.bib}

\end{document}